\documentclass[pre,twocolumn,twoside,byrevtex,superscriptaddress,showpacs,floatfix]{revtex4-1}

\usepackage [english]{babel}
\usepackage [autostyle, english = american]{csquotes}
\usepackage{indentfirst}
\MakeOuterQuote{"}

\usepackage{currfile}
\PassOptionsToPackage{hyphens}{url}\usepackage{hyperref}

\usepackage{graphicx,epsfig,verbatim,enumerate}
\usepackage{amssymb,amsmath}
\usepackage{ifthen}

\usepackage{longtable}

\usepackage{mathtools}

\newboolean{twocolswitch}

\newcommand{\avg}[1]{\left\langle#1\right\rangle}
\newcommand{\tavg}[1]{\langle#1\rangle}

\newcommand{\sindex}[1]{}
\newcommand{\nindex}[1]{}

\newcommand{\www}[1]{\url{#1}}

\newcommand{\Req}[1]{Eq.~(\ref{#1})}

\usepackage{lettrine}

\usepackage{color}

\newcommand{\dendrodepth}{d}

\newcommand{\gameratio}{R}

\newcommand{\leadsize}{\ell}

\setboolean{twocolswitch}{true}

\begin{document}

\title{\protect
The game story space of professional sports: Australian Rules Football
}

\author{
\firstname{Dilan Patrick}
\surname{Kiley}
}
\email{dilan.kiley@uvm.edu}

\affiliation{Department of Mathematics \& Statistics,
  Vermont Complex Systems Center,
  Computational Story Lab,
  \& the Vermont Advanced Computing Core,
  The University of Vermont,
  Burlington, VT 05401.}

\author{
\firstname{Andrew J.}
\surname{Reagan}
}
\email{andrew.reagan@uvm.edu}

\affiliation{Department of Mathematics \& Statistics,
  Vermont Complex Systems Center,
  Computational Story Lab,
  \& the Vermont Advanced Computing Core,
  The University of Vermont,
  Burlington, VT 05401.}

\author{
\firstname{Lewis}
\surname{Mitchell}
}
\email{lewis.mitchell@adelaide.edu.au}
\affiliation{
  School of Mathematical Sciences,
  North Terrace Campus,
  The University of Adelaide,
  SA 5005, Australia
}

\author{
\firstname{Christopher M.}
\surname{Danforth}
}
\email{chris.danforth@uvm.edu}

\affiliation{Department of Mathematics \& Statistics,
  Vermont Complex Systems Center,
  Computational Story Lab,
  \& the Vermont Advanced Computing Core,
  The University of Vermont,
  Burlington, VT 05401.}

\author{
\firstname{Peter Sheridan}
\surname{Dodds}
}
\email{peter.dodds@uvm.edu}

\affiliation{Department of Mathematics \& Statistics,
  Vermont Complex Systems Center,
  Computational Story Lab,
  \& the Vermont Advanced Computing Core,
  The University of Vermont,
  Burlington, VT 05401.}

\date{\today}

\begin{abstract}
  \protect
  Sports are spontaneous generators of stories.
Through skill and chance, 
the script of each game is dynamically written in real time
by players acting out possible trajectories allowed
by a sport's rules.
By properly characterizing a given sport's ecology of `game stories',
we are  able to capture the sport's capacity for unfolding interesting narratives, 
in part by contrasting them with random walks.
Here, we explore the game story space afforded 
by a data set of 1,310 Australian Football League (AFL) score lines.
We find that AFL games exhibit a continuous spectrum of stories rather
than distinct clusters.
We show how coarse-graining reveals identifiable motifs
ranging from last minute comeback wins to one-sided blowouts.
Through an extensive comparison with biased random walks, 
we show that real AFL games deliver a broader array of motifs than null models,
and we provide consequent insights into the narrative appeal of real games.

\end{abstract}

\pacs{89.65.-s, 89.20.-a, 05.40.Jc, 02.50.Ey}

\maketitle

\section{Introduction}
\label{sec:sog.intro}

While sports are often analogized to a wide array of other arenas
of human activity---particularly war---well known story lines and elements of sports are 
conversely invoked to describe other spheres.
Each game generates a probablistic, rule-based story~\cite{billings2013a},
and the stories of games provide a range of motifs which
map onto narratives found across the human experience:
dominant, one-sided performances; back-and-forth struggles; 
underdog upsets; and improbable comebacks.
As fans, people enjoy watching
suspenseful sporting events---unscripted stories---and 
following the fortunes of their favorite players and
teams~\cite{bryant1994a,gantz1995a,hugenberg2008a}. 

Despite the inherent story-telling nature of
sporting contests---and notwithstanding the vast statistical analyses
surrounding professional sports including
the many observations of
and departures from
randomness~\cite{elderton1945a,wood1945a,colwell1982a,bocskocsky2014a,ribeiro2012a,gabel2012a,clauset2015a}---the 
ecology of game stories
remains a largely unexplored, data-rich area~\cite{merritt2014a}.
We are interested in a number of basic questions such as 
whether the game stories of a sport form a spectrum or a set of relatively
isolated clusters, 
how well models such as random walks fare in reproducing the
specific shapes of real game stories,
whether or not these stories are compelling to fans, 
and how different sports compare in the
stories afforded by their various rule sets.

Here, we focus on
Australian Rules Football, a high skills game 
originating in the mid 1800s.
We describe Australian Rules Football in brief and then move
on to extracting and evaluating the sport's possible game stories.
Early on, the game evolved into a winter sport quite
distinct from other codes such as soccer or rugby while
bearing some similarity to Gaelic football.
Played as state-level competitions for most of the 1900s
with the Victorian Football League (VFL) being most prominent,
a national competition emerged in the 1980s
with the Australian Football League (AFL) becoming 
a formal entity in 1990.  
The AFL is currently constituted
by 18 teams located in five of Australia's states.

Games run over four quarters, each lasting around 30 minutes (including stoppage time),
and teams are each comprised of 18 on-field players.
Games (or matches) are played
on large ovals typically used for cricket in the summer
and of variable size (generally 135 to 185 meters in length).  
The ball is oblong and may be kicked
or handballed (an action where the ball is punched off one hand with the closed
fist of the other) but not thrown.
Marking (cleanly catching a kicked ball) is a central feature of the game,
and the AFL is well known for producing many spectacular marks and kicks for goals~\cite{jesaulenko1970a}.

The object of the sport is to kick goals, with the 
customary standard of highest score wins (ties are relatively rare but possible).
Scores may be 6 points or 1 point as follows, some minor details aside.
Each end of the ground has four tall posts.  Kicking the ball
(untouched) through the central two posts results in a `goal' or 6
points.  If the ball is touched or goes through either of the outer 
two sets of posts, then the score is a `behind' or 1 point.
Final scores are thus a combination of goals (6) and behinds (1)
and on average tally around 100 per team.  Poor conditions or poor play may lead to 
scores below 50, while scores above 200 are achievable in the case
of a `thrashing' (the record high and low scores are 239 and 1).
Wins are worth 4 points, ties 2 points, and losses 0.

\begin{figure}[tbp!]
  \centering
  \includegraphics[width=\columnwidth]{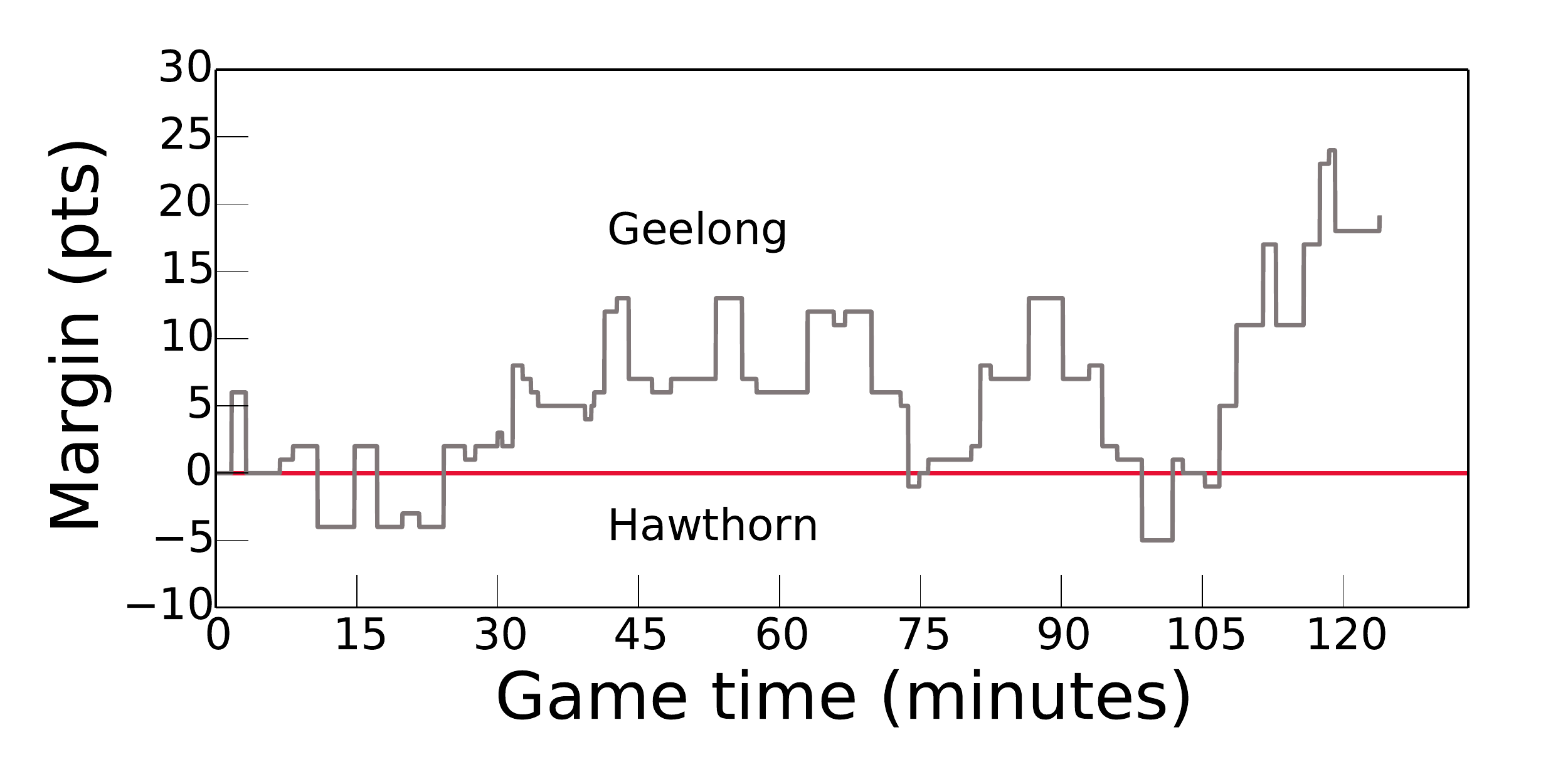}
  \caption{ 
    Representative `game story' (or score differential `worm') for
    an example AFL contest held between
    Geelong and Hawthorn on Monday April 21, 2014. 
    Individual scores are either goals (6 points) or behinds (1 point).
    Geelong won by 19 with a final score line of
    106 (15 goals, 16 behinds) to 87 (12 goals, 15 behinds).
  }
  \label{fig:sog.example_worm}
\end{figure}

Of interest to us here is that the AFL provides an excellent test case for 
extracting and describing the game story space of a professional sport.
We downloaded 1,310 AFL game scoring
progressions from
\href{http://www.afltables.com}{http://afltables.com} 
(ranging from the 2008 season to midway through the 2014 season)~\cite{afl_tables_www.afltables.com_2014}. 
We extracted the scoring dynamics of each game down to
second level resolution, with the possible events
at each second being 
(1) a goal for either team, 
(2) a behind for either team, 
or 
(3) no score~\cite{australian_football_league_laws_2015}. 
Each game thus affords a `worm' tracking the score differential between two
teams.  We will call these worms `game stories' and
we provide an example in Fig.~\ref{fig:sog.example_worm}.
The game story shows that Geelong pulled away from 
Hawthorn---their great rival over the preceding decade---towards the end of
a close, back and forth game.

Each game story provides a rich representation of a game's flow, and,
at a glance, quickly indicates key aspects such as largest lead,
number of lead changes, momentum swings, and one-sidedness. 
And game stories evidently allow for a straightforward 
quantitative comparison between
any pair of matches.

For the game story ecology we study here,
an important aspect of the AFL is that rankings (referred to as the ladder), 
depend first on number of wins (and ties), and then percentage of
`points for' versus `points against'.
Teams are therefore generally motivated to score as heavily as
possible
while still factoring in increased potential for injury.

We order the paper as follows.
In Sec.~\ref{sec:sog.basics},
we first present a series of basic
observations about the statistics of AFL games.
We include an analysis of conditional
probabilities for winning as a function of lead size.
We show through a general comparison to random walks 
that AFL games are collectively more diffusive
than simple random walks leading to a biased random
walk null model based on skill differential between teams.
We then introduce an ensemble of 100 sets of 1,310 biased random walk game stories
which we use throughout the remainder of the paper.
In Secs.~\ref{sec:sog.gameshapes}
and \ref{sec:sog.gamemotifs},
we demonstrate that game stories form a spectrum rather than
distinct clusters,
and we apply coarse-graining to elucidate game story motifs at two levels of resolution.
We then provide a detailed comparison between real game motifs 
and the smaller taxonomy of motifs 
generated 
by our biased random walk null model.
We explore the possibility of predicting final game margins in 
Sec.~\ref{sec:sog.prediction}. 
We offer closing thoughts and propose
further avenues of analysis in Sec.~\ref{sec:sog.conclusion}.

% We  provide visualizations of story motifs 

\section{Basic game features}
\label{sec:sog.basics}

\subsection{Game length}
\label{subsec:sog.gamelength}

While every AFL game is officially comprised of four 20 minute quarters of
playing time, the inclusion of stoppage time means there is no set 
quarter or game length, resulting in some minor complications for our analysis.
We see an approximate Gaussian distribution of game lengths with the average game lasting a
little over two hours at 122 minutes,
and 96\% of games run for around 112 to
132 minutes ($\sigma \simeq 4.8$ minutes).
In comparing AFL games, we must 
therefore accommodate different game lengths.
A range of possible approaches include dilation, truncation, and
extension (by holding a final score constant), and we will explain and
argue for the latter
in Sec.~\ref{sec:sog.gameshapes}.

\subsection{Scoring across quarters}
\label{subsec:sog.scoring-quarters}

In post-game discussions, commentators will often focus on the natural
chapters of a given sport.
For quarter-based games, matches will sometimes be
described as `a game of quarters' or `a tale of two halves.'
For the AFL, we find that scoring does not, on average, vary greatly as the game progresses from
quarter to quarter (we will however observe interesting quarter-scale motifs later on).
For our game database, we find there is slightly more scoring done in the second
half of the game (46.96 versus 44.91), 
where teams score
one more point, on average, in the fourth quarter versus the first quarter
(23.48 versus 22.22).
This minor increase may be due to a heightened sense of the importance of each point
as game time begins to run out,
the fatiguing of defensive players,
or as a consequence of having `learned an opponent'~\cite{merritt2014a,thompson2008a}.

\begin{figure}[tbp!]
  \centering
  \includegraphics[width=\columnwidth]{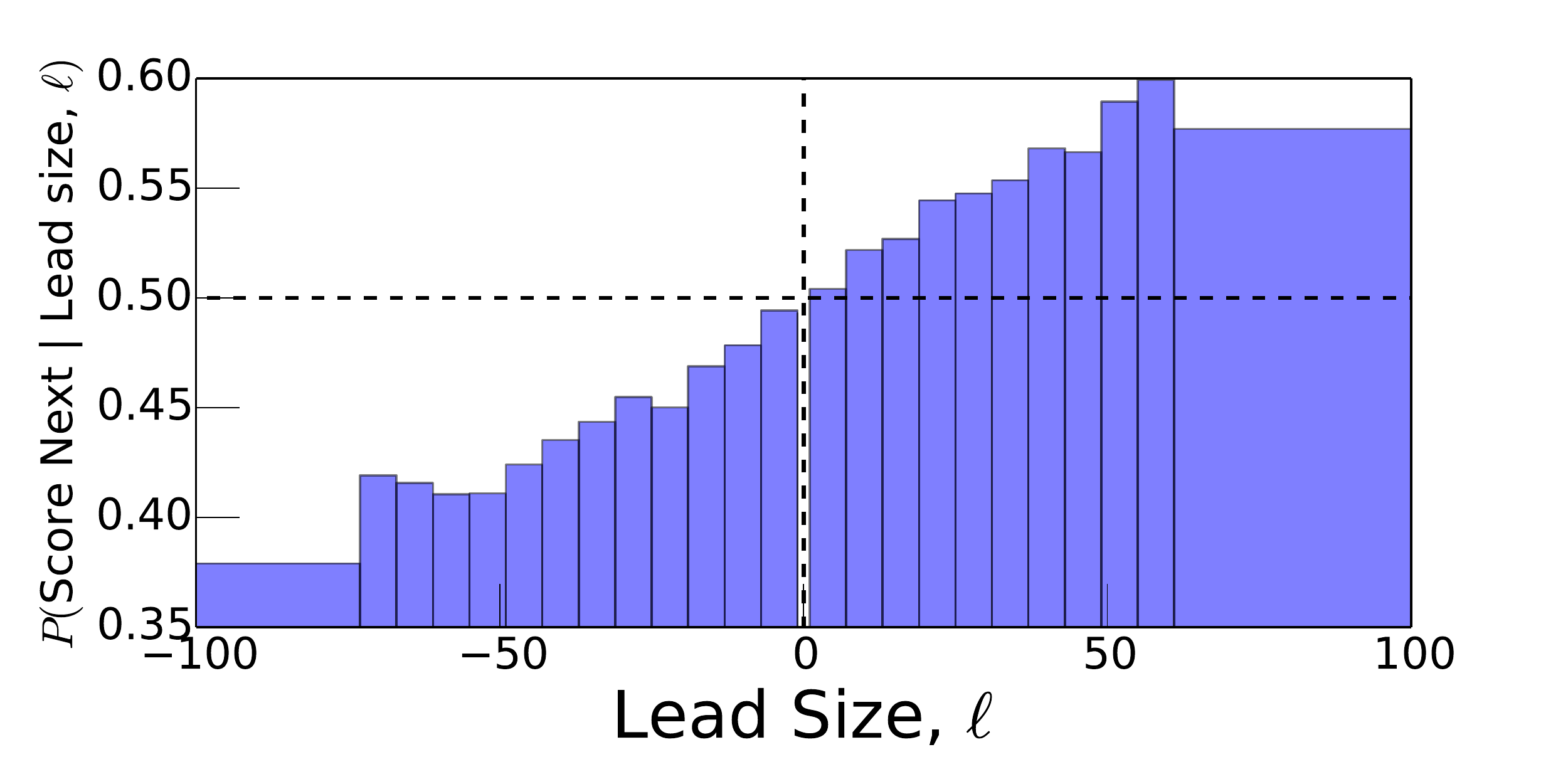}
  \caption{ 
    Conditional probability of scoring the next goal or behind given a particular lead
    size. 
    Bins are in six point blocks with the extreme leads collapsed: 
    $<$ -72, -71 to -66, \ldots, -6 to -1, 1 to 6, 7 to 12, \ldots, $>$ 72.
    As for most sports,
    the probability of scoring next
    increases approximately linearly as a function of current lead size.
  }
  \label{fig:sog.conditional_scoring}
\end{figure}

\subsection{Probability of next score as a function of lead size}
\label{subsec:sog.nextscore}

In Fig.~\ref{fig:sog.conditional_scoring}, we show that,
as for a number of other sports, the probability of scoring next
(either a goal or behind) 
at any point in a game
increases linearly as a function of the current lead size
(the National Basketball Association is a clear exception)~\cite{berger2011a,gabel2012a,merritt2014a,clauset2015a}.
This reflects a kind of momentum gain within games,
and could be captured by a simple biased model 
with scoring probability linearly tied to the current lead.
Other studies have proposed this linearity to be the result of a heterogeneous skill
model~\cite{merritt2014a}, and, as we describe in the following
section, we use a modification of such an approach.

\subsection{Conditional probabilities for winning}
\label{subsec:sog.winning}

\begin{figure}[tbp!]
  \centering
  \includegraphics[width=0.49\textwidth]{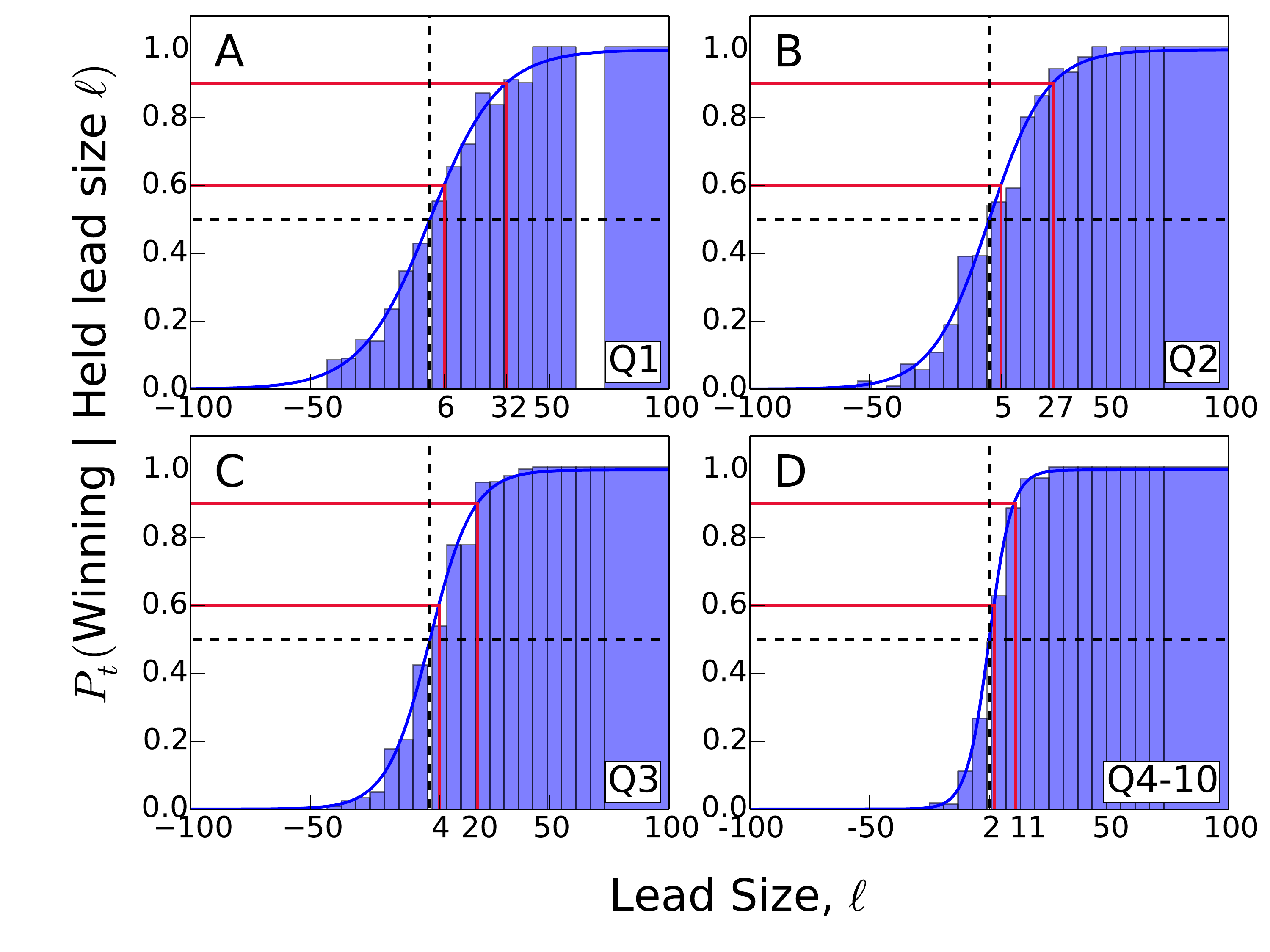}
  \caption{ 
    Conditional probability of winning given a lead of size 
    $\leadsize$
    at the end of the first three quarters 
    \textbf{(A--C)} 
    and with
    10 minutes to go in the game
    \textbf{(D)}.
    Bins are 
    comprised of the aggregate of every 6 points as in 
    Fig.~\ref{fig:sog.conditional_scoring}. 
    The dark blue curve is a sigmoid function of the form
    $
    [
    1 
    + 
    e^{-k(\leadsize-\leadsize_0)}
    ]^{-1}
    $
    where $k$ and $\leadsize_{0}$ are fit parameters
    determined via standard optimization 
    using the Python function scipy.optimize.curve\_fit
    (Note that $\leadsize_{0}$ should be 0 by construction.)
    As a game progresses, the threshold for
    likely victory (winning probability 0.90, upper red lines) decreases as
    expected, as does a threshold for a close game (probability of
    0.60, lower red line).
    The slope of the sigmoid curve increases as the game time progresses
    showing the evident greater impact of each point.
    We note that the missing data in panel \textbf{A} is a real feature
    of the specific 1,310 games in our data set.
  }
  \label{fig:sog.conditional_winning_quarters}
\end{figure}

We next examine the conditional probability 
of winning  given a lead of size $\leadsize$
at a time point $t$ in a game,
$P_{t}(\textnormal{Winning}\,|\,\leadsize) $.
We consider four example time points---the end of each of the first three quarters and with 10 minutes 
left in game time---and plot the results in Fig.~\ref{fig:sog.conditional_winning_quarters}.
We fit a sigmoid curve (see caption) to each conditional probability.
As expected, we immediately see an increase in winning probability for
a fixed lead as the game progresses.

These curves could be referenced
to give a rough indication of an unfolding game's likely outcome
and may be used to generate a range of statistics.
As an example,
we define likely victory as $P(\textnormal{Winning}\,|\,\leadsize) \ge 0.90$ and 
find $\leadsize$ = 32, 27, 20, and 11 are the approximate 
corresponding lead sizes
at the four time points.  Losing games after holding any
of these leads might be viewed as `snatching defeat from
the jaws of victory.'

Similarly, if we define close games as those
with $P(\textnormal{Winning}\,|\,\leadsize) \le 0.60$, we find the corresponding
approximate lead sizes to be $\leadsize \simeq$ 6, 5, 4, and 2.
These leads could function in the same way as the save
statistic in baseball is used, i.e., to acknowledge when a pitcher performs
well enough in a close game to help ensure their team's victory.
Expanding beyond the AFL, such probability thresholds for likely
victory or uncertain outcome may be modified to apply
to any sport, and could be greatly refined using detailed
information such as recent performances, stage of a season, and
weather conditions.

\section{Random walk null models}
\label{sec:sog.randomwalks}

A natural null model for a game story
is the classic, possibly biased, random walk~\cite{feller1971a,gabel2012a}. 
We consider an ensemble of modified random walks,
with each walk 
(1) composed of steps of $\pm$ 6 and $\pm$ 1,
(2) dictated by a randomly drawn bias,
(3) running for a variable total number of events,
and 
(4) with variable gaps between events, all informed by
real AFL game data.
For the purpose of exploring motifs later on, 
we will create 100 sets of 1,310 games.

An important and subtle aspect of the null model is the 
scoring bias, which we will denote by $\rho$.
We take the bias for each game simulation to be a proxy
for the skill differential between two opposing teams, 
as in \cite{merritt2014a},
though our approach involves an important adjustment.

In~\cite{merritt2014a}, a symmetric skill bias distribution 
is generated by taking the relative number of scoring events
made by one team in each game.
For example, given a match between two teams $T_1$ and $T_2$,
we find  
the number of scoring events generated by $T_{1}$, $n_{1}$,
and 
the same for $T_{2}$, $n_{2}$.
We then estimate a posteriori the skill bias between
the two teams as:
\begin{equation}
  \rho 
  = 
  \frac{n_{1}}
  {n_{1}+n_{2}}.
  \label{sog.eq:balance}
\end{equation}
In constructing the distribution of $\rho$, $f(\rho)$,
we discard information regarding how specific teams perform against
each other over seasons and years, and we are thus
only able to assign skill bias in a random, memoryless fashion
for our simulations.
We also note that for games with more than one value
of points available for different scoring events (as in 6 and 1 for
Australian Rules Football),
the winning team may register less scoring events than the losing one.

\begin{figure}[tbp!]  
  \centering
  \includegraphics[width=\columnwidth]{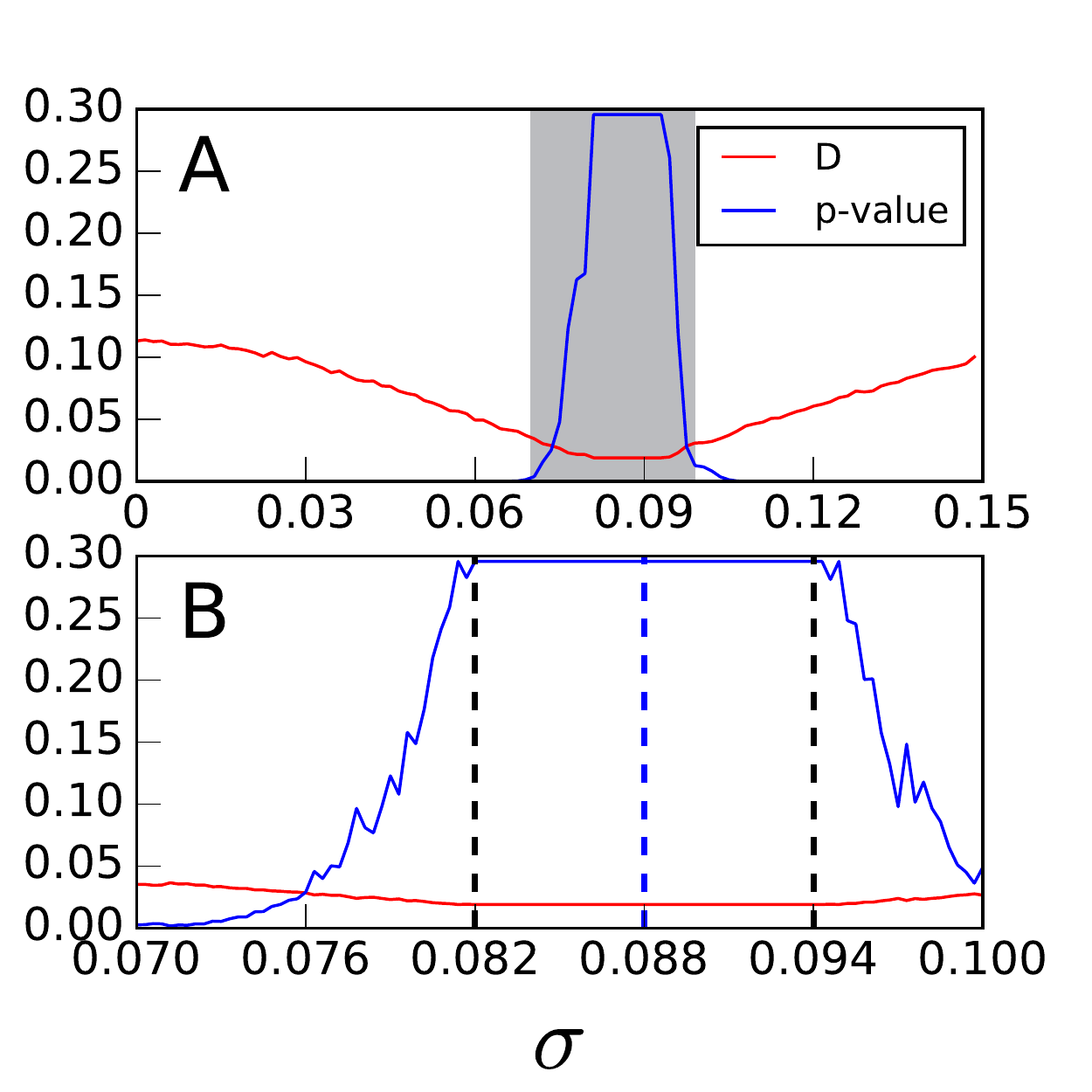}
  \caption{
    Skill bias $\rho$ represents a team's relative ability to score against
    another team and is estimated a posteriori by
    the fraction of scoring events made by each team~\Req{sog.eq:balance}.
    \textbf{A} and \textbf{B}: Kolmogorov-Smirnov test $D$ statistic and associated $p$-value
    comparing the observed output skill bias distribution 
    produced by a presumed input skill distribution
    $f$ with that observed 
    for all AFL games in our data set,
    where $f$ is Gaussian with mean $0.5$ and its standard
    deviation $\sigma$ is the variable of interest.
    For each value of $\sigma$, we created 1,000 
    biased random walks with the bias $\rho$ drawn from the
    corresponding normal distribution.
    Each game's number of events was 
    drawn from a distribution of the number of events in real
    AFL games (see text).
    Plot \textbf{B} is an expanded version of the shaded region
    in \textbf{A} with finer sampling.
    We estimated the best fit
    to be $\sigma \simeq 0.088$,
    and we compare the resulting observed bias distribution
    with that of~\cite{merritt2014a}
    in Fig.~\ref{fig:sog.Resulting_Event_Biases}.
  }
  \label{fig:sog.Sigma_Fitting}
\end{figure}

In~\cite{merritt2014a}, random walk game stories were then
generated directly using $f(\rho)$.
However, for small time scales this is immediately problematic and requires
a correction.
Consider using such an approach on pure random walks.
We of course have that $f(\rho) = \delta(\rho-1/2)$ by construction,
but our estimate of $f(\rho)$
will be a Gaussian of width $\sim t^{-1/2}$, where we have
normalized displacement by time $t$.
And while as $t \rightarrow \infty$, our estimate of $f(\rho)$ approaches
the correct distribution $\delta(\rho-1/2)$, 
we are here dealing with relatively short random walks.
Indeed, we observe that if we start with pure random walks,
run them for, say, 100 steps, estimate the bias distribution,
run a new set of random walks with these biases, and keep repeating
this process, we obtain an increasingly flat bias distribution.

To account for this overestimate of the spread of skill bias,
we propose the tuning of an input Gaussian distribution of skill biases 
so as to produce biased random walks whose outcomes best 
match the observed  event biases for real games.
We assume that $f$ should be centered at $\rho = 0.50$. 
We then draw from an appropriate distribution of number of events per
game,  
and tune the standard deviation of $f$, $\sigma$, to minimize the 
Kolmogorov-Smirnov (KS) $D$ statistic and maximize
the $p$-value produced from a two-tailed KS test between the resulting
distribution of event biases and the underlying, observed distribution
for our AFL data set.

\begin{figure}[tbp!]
  \centering
  \includegraphics[width=\columnwidth]{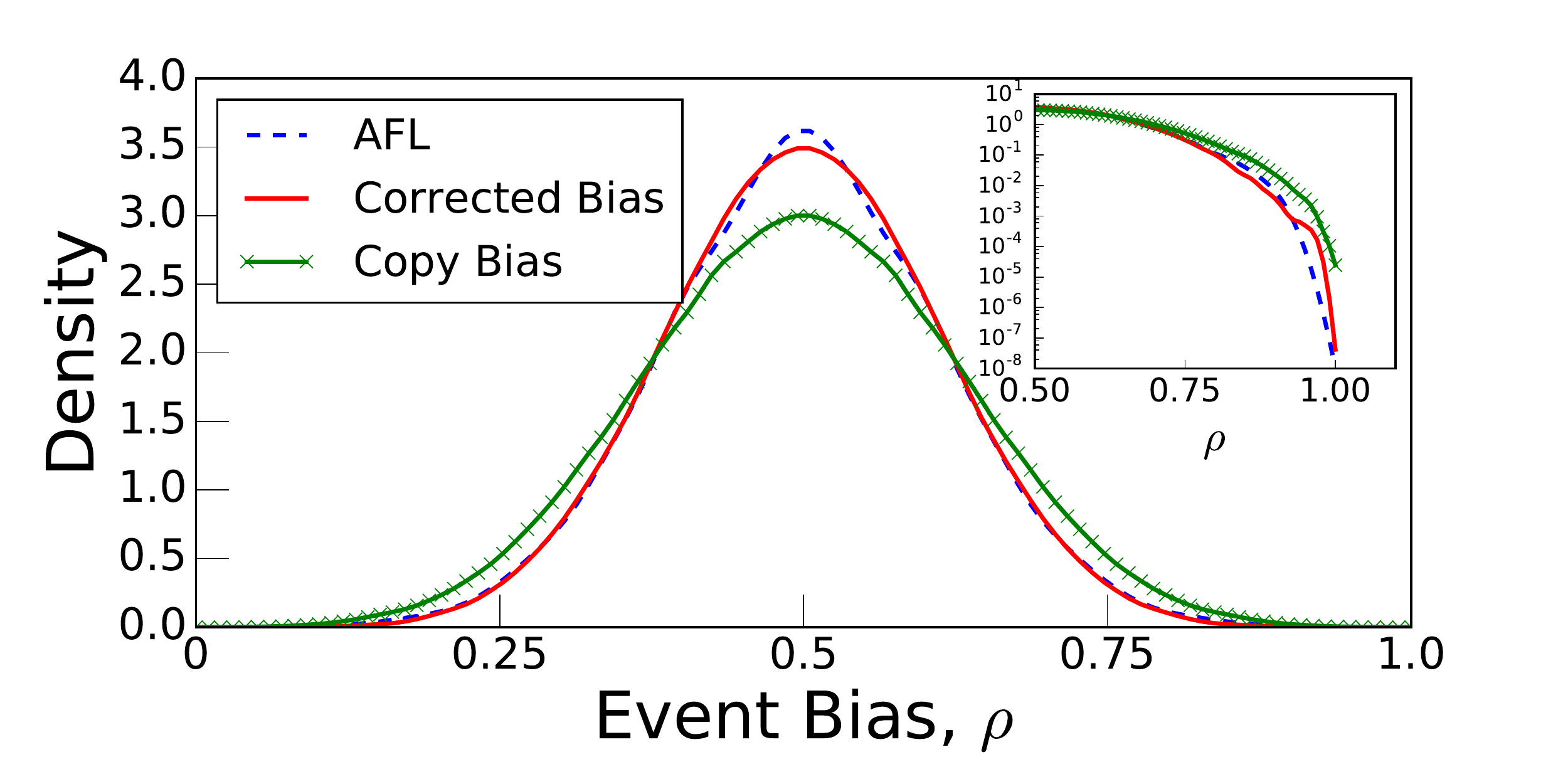}
  \caption{ 
    Comparison of the observed AFL skill bias distribution 
    (balance of scoring events $\rho$ given in \Req{sog.eq:balance}, dashed blue curve)
    with that produced by two approaches:
    (1) We draw $\rho$ from a normal distribution using the best
    candidate $\sigma$ value with mean $0.50$ as determined 
    via Fig.~\ref{fig:sog.Sigma_Fitting} (red curve), 
    and
    (2) We choose $\rho$ 
    from the complete list of observed biases from the AFL 
    (green curve, the replication method of \cite{merritt2014a}).
    For the real and the two simulated distributions,
    both $\rho$ and $1-\rho$ are included
    for symmetry.
    The fitted $\sigma$ approach produces a more accurate
    estimate of the observed biases, particularly for 
    competitive matches ($\rho$ close to 0.50) and
    one sided affairs.
    Inset: Upper half of the distributions plotted on a
    semi-logarithmic scale (base 10)
    revealing that the replication method of \cite{merritt2014a}
    also over produces extreme
    biases, as compared to the AFL and our proposed correction using a
    numerically determined $\sigma$.
  }
  \label{fig:sog.Resulting_Event_Biases}
\end{figure}

We show the variation of $D$ and the $p$-value as a function of
$\sigma$ in Fig.~\ref{fig:sog.Sigma_Fitting}.
We then demonstrate in Fig.~\ref{fig:sog.Resulting_Event_Biases}
that the $\sigma$-corrected distribution produces an observably better
approximation of outcomes than if we used the observed biases approach
of~\cite{merritt2014a}.
Because the fit for our method in
Fig.~\ref{fig:sog.Resulting_Event_Biases}
is not exact,
a further improvement (unnecessary here) would be to 
allow $f$ to be arbitrary rather than assuming a Gaussian.

With a reasonable estimate of $f$ in hand, 
we create 100 ensembles of 1,310 null games where each game is
generated with 
(1) one team scoring with probability $\rho$ drawn from
the $\sigma$-corrected distribution described above; 
(2) individual
scores being a goal or behind with probabilities based on the AFL
data set (approximately 0.53 and 0.47); and 
(3) a variable number of
events per simulation based on: 
(a) game duration drawn from the
approximated normal distribution described in
Sec.~\ref{sec:sog.basics}, 
and (b) time between events drawn from a
Chi-squared distribution fit to the inter-event times of real games.

\begin{figure}[tbp!]
  \centering
  \includegraphics[width=\columnwidth]{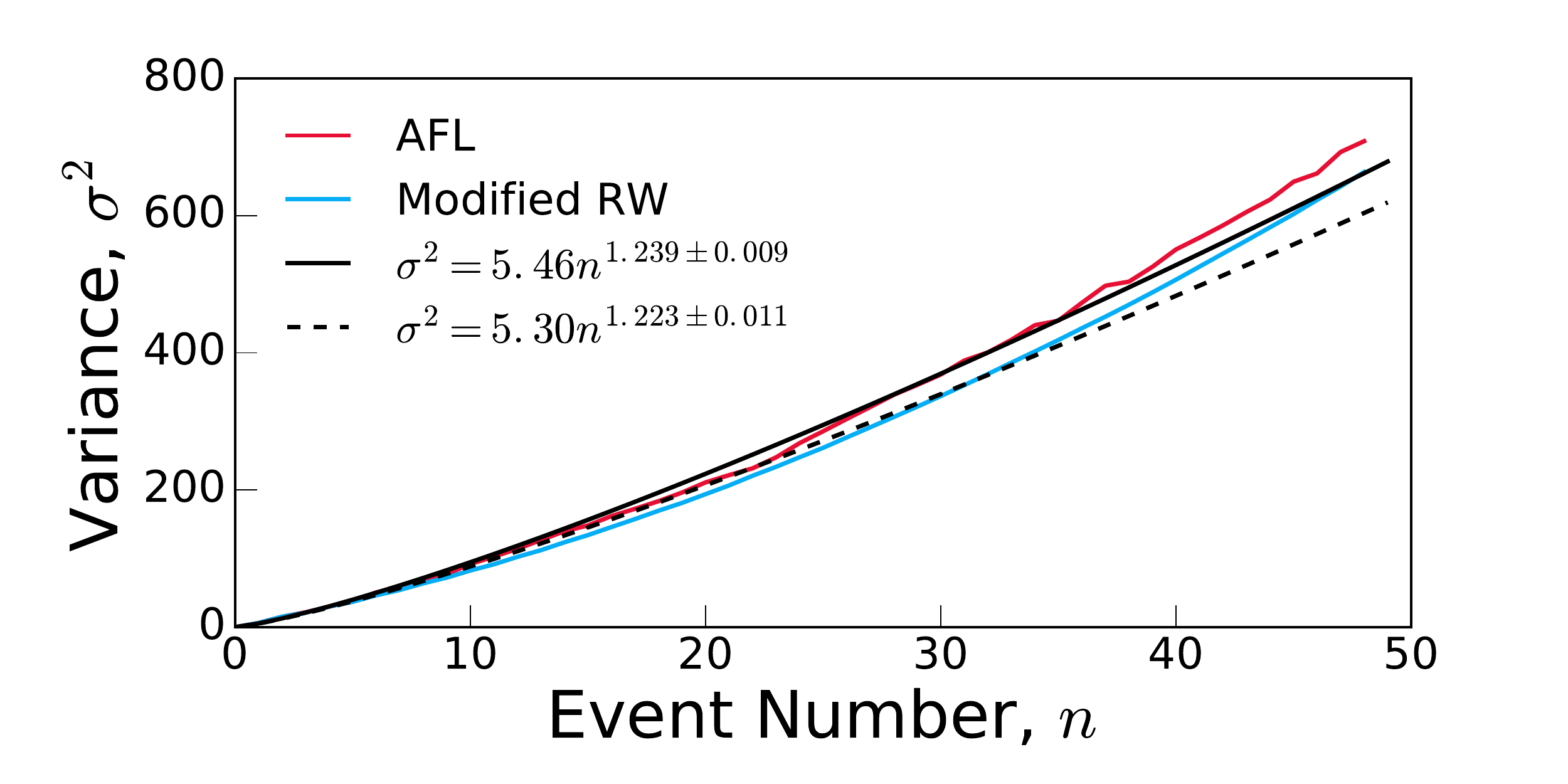}
  \caption{
    Variance in the instantaneous margin
    as a function of event number for
    real AFL games
    (solid red curve)
    and biased random walks
    as described in Sec.~\ref{sec:sog.randomwalks},
    (solid blue curve).
    We perform fits in logarithmic space using standard
    least squares regression 
    (solid black curve for real games, 
    dashed black
    for the null model).
    The biased random walks 
    satisfactorily reproduce the observed scaling of variance.
    It thus appears that AFL games stories do not exhibit inherently
    superdiffusive behavior but rather result from imbalances between
    opposing teams.
  }
  \label{fig:sog.biased_margin_variance}
\end{figure}

For a secondary test on the validity of our null model's game stories,
we compute the variance $\sigma^{2}$
of the margin at each event number $n$
for both AFL games and modified random walks 
(for the AFL games, we orient each walk according
to home and away status, the default ordering in the data set).
As we show in Fig.~\ref{fig:sog.biased_margin_variance},
we find that both AFL games and biased random walks 
produce game stories with 
$
\sigma^2
\sim 
n^{1.239 \pm 0.009}
$
and
$
\sigma^2
\sim 
n^{1.236 \pm 0.012}
$
respectively.
Collectively, AFL games thus have a tendency toward runaway score
differentials, 
and while superdiffusive-like, this superlinear scaling 
of the variance can be almost entirely accounted for by our incorporation
of the skill bias distribution $f$.

\section{Measuring distances between games}
\label{sec:sog.gameshapes}

\begin{figure}[tbp!]
  \centering
  \includegraphics[width=.8\columnwidth]{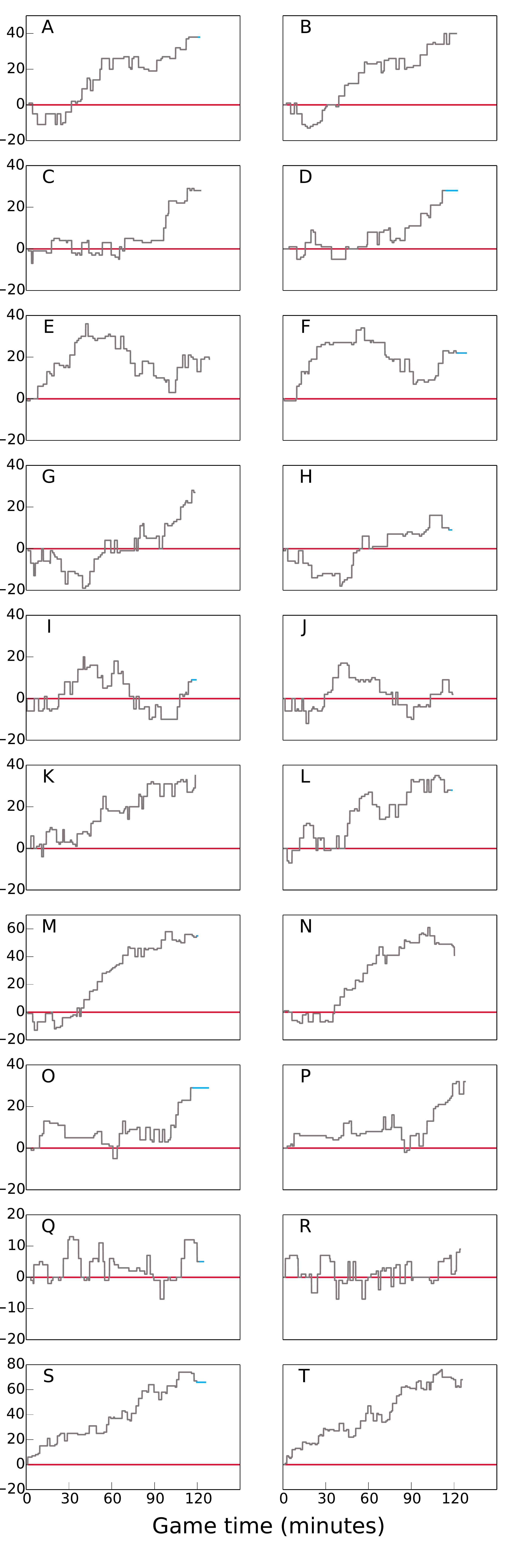}
  \caption{ 
    Top ten pairwise neighbors as determined by the distance
    measure between each game described by~\Req{eq:sog.gamedist}. 
    In all examples, dark gray curves denote the game story.
    For the shorter game of each pair, horizontal solid blue lines show
    how we hold the final score constant
    to equalize lengths of games.
  }
  \label{fig:sog.closest_games}
\end{figure}

Before moving on to our main focus, the ecology of game stories,
we define a straightforward measure of the distance between any pair of games.
For any sport,
we define a distance measure between
two games $i$ and $j$ as
\begin{equation}
D(g_{i},g_{j}) 
= 
T^{-1}
\sum_{t=1}^{T} 
\left| 
  g_{i}(t) - g_{j}(t)
\right|,
\label{eq:sog.gamedist}
\end{equation}
where $T$ is the length of the game in seconds, and $g_{i}(t)$ is the
score differential between the competing teams in game $i$ at second $t$. 
We orient game stories so that the team whose score is 
oriented upwards on the vertical axis wins or ties
[i.e., $g_i(T) \ge 0$]. 
By construction, pairs of games which have a relatively small distance
between them will have similar game stories.
The normalization factor $1/T$ means the distance remains in the units of points and
can be thought of as the average difference between point
differentials over the course of the two games. 

In the case of the AFL, due to the fact that games do not
run for a standardized time $T$,
we extend the game story of the shorter of the pair 
to match the length of the longer game by holding the final score constant.
While not ideal, we observe that the metric performs well in identifying games that are closely related. 
We investigated several alternatives such as linearly dilating the
shorter game, and found no compelling benefits.  Dilation may
be useful in other settings but the distortion of real time
is problematic for sports.

In Fig.~\ref{fig:sog.closest_games}, we present the ten most similar
pairs of games in terms of their stories.
These close pairs show the metric performs as it should
and that, moreover, proximal games are not dominated by a certain type. 
Figs.~\ref{fig:sog.closest_games}A and \ref{fig:sog.closest_games}B demonstrate a team
overcoming an early stumble, 
Figs.~\ref{fig:sog.closest_games}E and \ref{fig:sog.closest_games}F showcase the victor repelling an
attempted comeback, 
Figs.~\ref{fig:sog.closest_games}Q and \ref{fig:sog.closest_games}R exemplify a see-saw battle with many lead
changes, 
and 
Fig.~\ref{fig:sog.closest_games}S and \ref{fig:sog.closest_games}T capture blowouts---one team taking control early and
continuing to dominate the contest.

\section{Game Story Ecology}
\label{sec:sog.gamemotifs}

Having described and implemented a suitable metric for comparing
games and their root story, 
we seek to group games together with the objective
of revealing large scale characteristic motifs. 
To what extent are well-known game narratives---from blowouts to
nail-biters to improbable comebacks---and potentially less well known story lines
featured in our collection of games?
And how does the distribution of real game stories compare with those
of our biased random walk null model?  
(We note that in an earlier
version
of the present paper, we considered pure, unbiased random walks for
the null model~\cite{kiley2015a}.)

\begin{figure}[tbp!]
  \centering
  \includegraphics[width=\columnwidth]{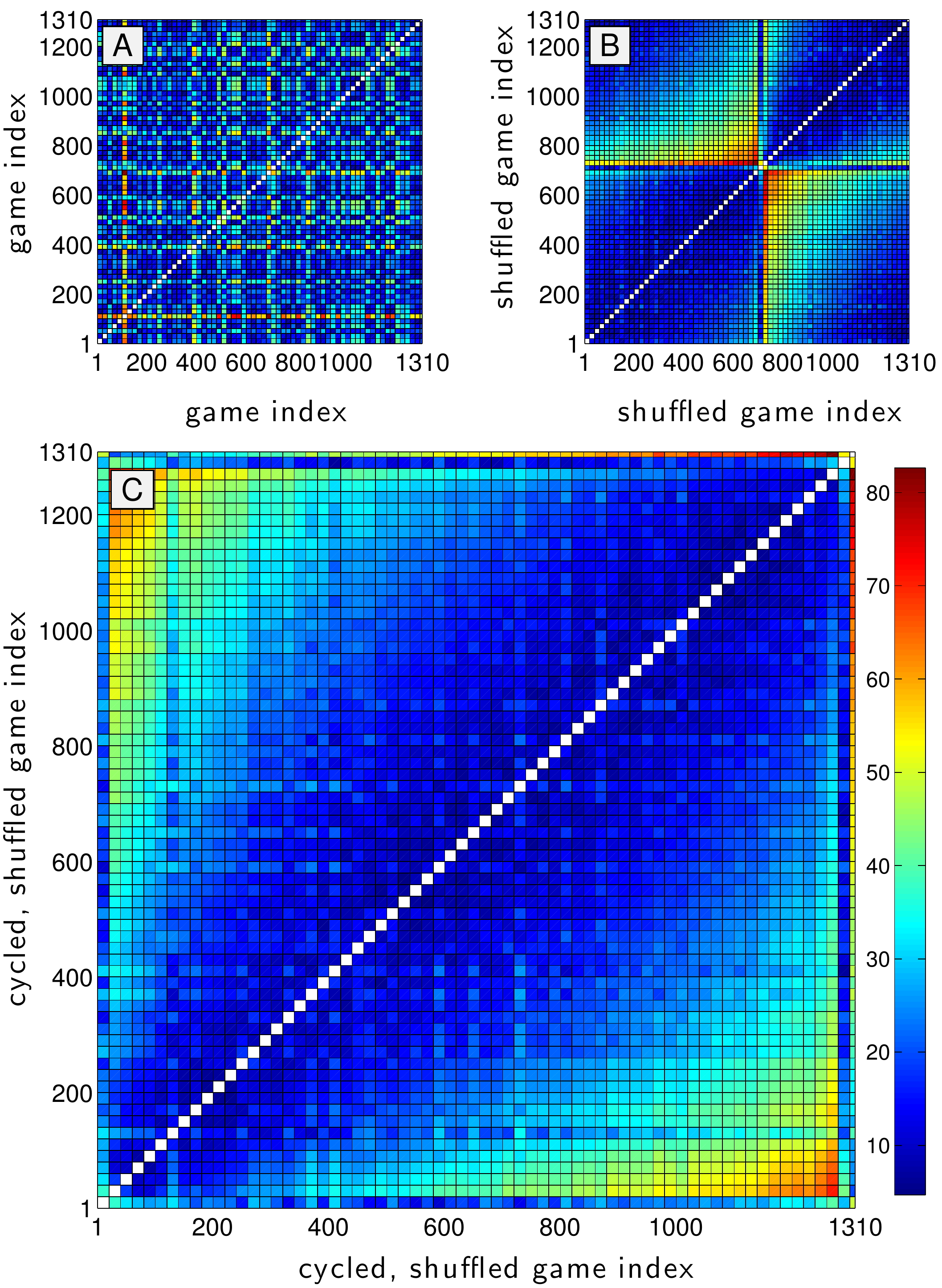}
  \caption{ 
    Heat maps for 
    \textbf{(A)}
    the pairwise distances between games unsorted on a ring;
    \textbf{(B)}
    the same distances after
    games have been reordered on the ring so as to minimize the
    cost function given in~\Req{eq:sog.ringcost};
    \textbf{(C)}
    the same as \textbf{(B)} 
    but with
    game indices cycled to make the continuous
    spectrum of games evident.  
    We include only every 20th game for clarity
    and note that such shuffling is usually
    performed for entities on a line rather than a ring.
    The games at the end of the spectrum
    are most dissimilar and correspond to runaway victories and
    comebacks
    (see also Fig.~\ref{fig:sog.dendro_heat}).
  }
  \label{fig:sog.minimize_by_shuffling_ring100_005}
\end{figure}

\begin{figure}[tbp!]
  \centering
  \includegraphics[width=\columnwidth]{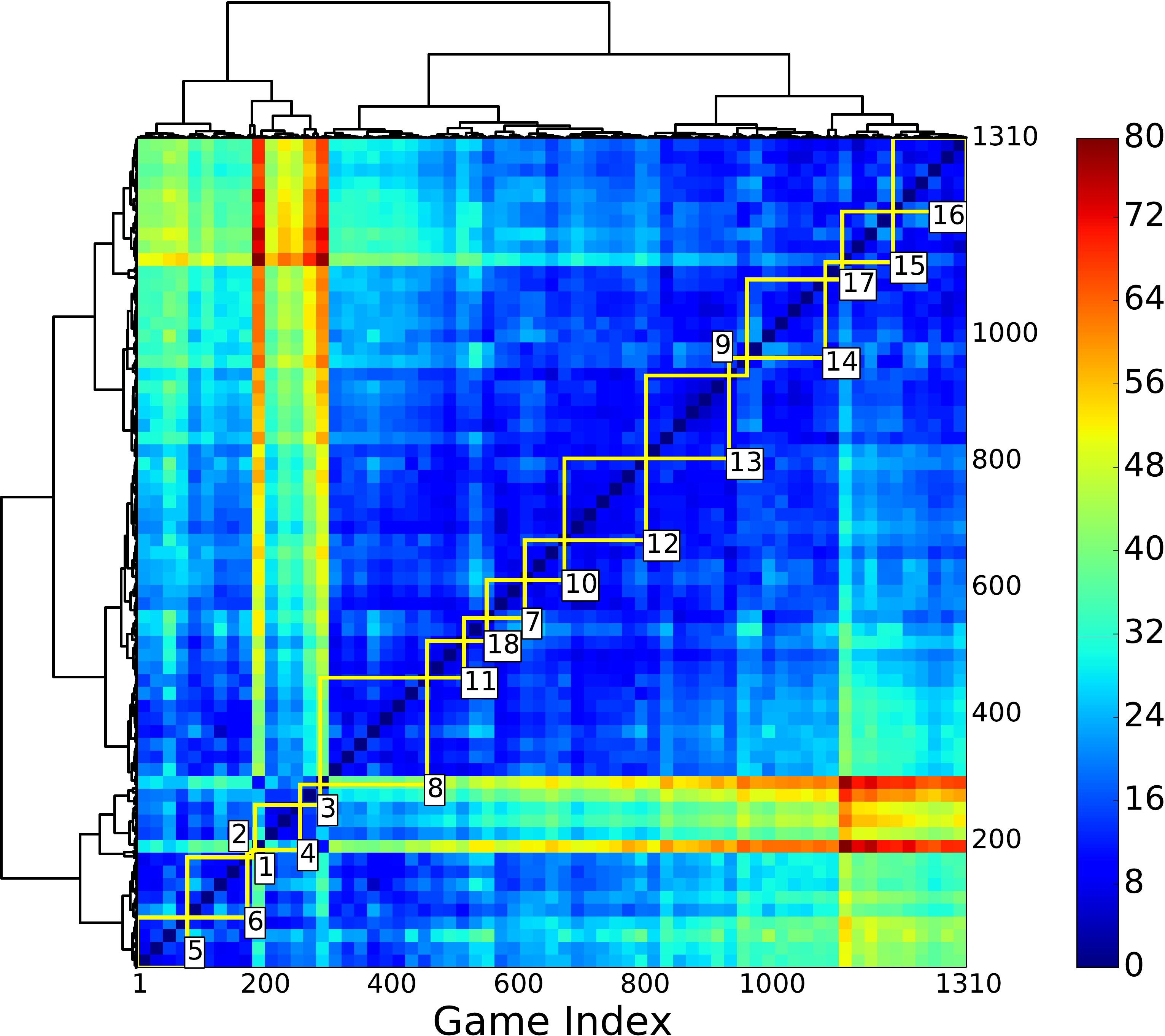}
  \caption{ 
    Heat matrix for the pairwise distances between games,
    subsampled by a factor of 20 as 
    per Fig.~\ref{fig:sog.minimize_by_shuffling_ring100_005}.
    A noticeable
    split is visible between the blowout games (first six clusters)
    and the comeback victories (last three clusters).
    We plot
    dendrograms along both the top and left edges of the
    matrix, and
    as explained in Sec.~\ref{subsec:18motifs},
    the boxed numbers reference the 18 motifs found
    when the average intra-cluster distance is set to 11 points.
    These 18 motifs are variously displayed in
    Figs.~\ref{fig:sog.biased18motifs}
    and 
    \ref{fig:sog.realmotifs18ratio-random}.
  }
  \label{fig:sog.dendro_heat}
\end{figure}

\subsection{AFL games constitute a single spectrum}
\label{subsec:sog.spectrum}

We first compute the pairwise distance between all games in our
data set.  
We then apply a shuffling algorithm 
to order games on a discretized ring so that similar games
are as close to each other as possible.
Specifically, we minimize the cost
\begin{equation}
  C 
  = 
  \sum_{i,j \in N, i \ne j} 
  d_{ij}^{\,2} 
  \cdot
  D(g_i,g_j)^{-1}
\label{eq:sog.ringcost}
\end{equation}
where $d_{ij}$ is the shortest distance between $i$ and $j$ on the
ring.
At each step of our minimization procedure,
we randomly choose a game and determine which
swap with another game most reduces $C$.
We use $d_{ij}^{\,2}$ by choice and other powers give similar results.

In Fig.~\ref{fig:sog.minimize_by_shuffling_ring100_005}, we show three heat maps for
distance $D$ with:
(A) games unsorted;
(B) games sorted according to the above minimization procedure;
and
(C) indices of sorted games cycled to reveal
that AFL games broadly constitute a continuous spectrum.
As we show below, at the ends of the spectrum are the most extreme
blow outs, 
and the strongest comebacks---i.e., one team dominates for the first half
and then the tables are flipped in the second half.

\subsection{Coarse-grained motifs}
\label{subsec:sog.coarsegraining}

While little modularity is apparent---there are no evident distinct
classes of games---we may nevertheless perform a kind 
of coarse-graining via hierarchical clustering
to extract a dendrogram of increasingly resolved game motifs.

Even though we have just shown that the game story ecology forms
a continuum, it is important that we stress that the motifs we find
should not be interpreted as well separated clusters.
Adjacent motifs will have similar game stories at their connecting 
borders.
A physical example might be the landscape roughness of equal 
area regions dividing up a country---two connected areas would typically be locally similar
along their borders.
Having identified a continuum, we are simply now addressing the variation
within that continuum using a range of scales.

\begin{figure}[tbp!]
  \centering
  \includegraphics[width=\columnwidth]{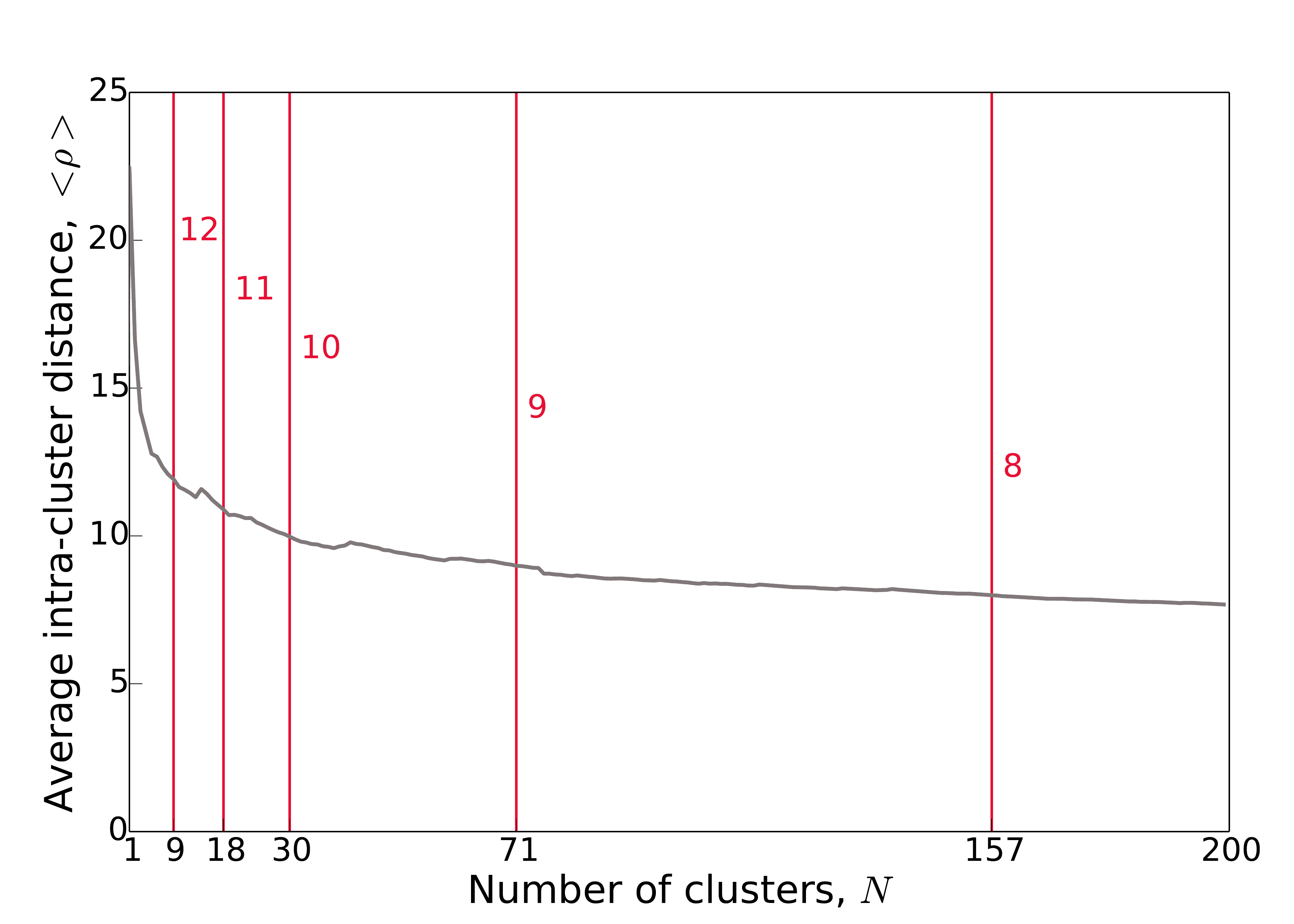}
  \caption{
    Average intra-cluster distance $\tavg{\rho}$ as a function of cluster number
    $N$. 
    Red lines mark the first occurrence in which the average of the intra
    cluster distance of the N motif clusters had a value below 12, 11,
    10, 9, and 8 (red text beside each line) points respectively.
    The next cut for 7 points gives 343 motifs.
  }
  \label{fig:sog.point_breaks}
\end{figure}

We employ a principled approach to
identifying meaningful levels of coarse-graining, leading to families of motifs.
As points are the smallest scoring unit in AFL games,
we use them to mark resolution scales as follows.
First, we define $\rho_{i}$, 
the average distance between
games within a given cluster $i$ as
\begin{equation}
\rho_{i}
= 
\frac{1}{
  n_i(n_i-1)
}
\sum_{j=1}^{n_i}
\sum_{k=1,k \ne j}^{n_i}
D(g_j,g_k).
\label{eq:sog.intraclusterdist}
\end{equation}
Here $j$ and $k$ are games placed in cluster $i$, 
$n_i$ is the number of games in cluster $i$, 
and $D$ is the game distance defined
in~\Req{eq:sog.gamedist}. 
At a given depth $\dendrodepth$ of the dendrogram, 
we compute $\rho_{i}(d)$ for each of
the $N(d)$ clusters found, 
and then average
over all clusters to obtain an
average intra-cluster distance:
\begin{equation}
\avg{\rho(d)}
= 
\frac{1}{N(d)}
\sum_{i=1}^{N(d)}
\rho(d).
\label{eq:sog.avgintraclusterdist}
\end{equation}

We use Ward's method of variance to construct a
dendrogram~\cite{ward1963a}, as shown in
Fig.~\ref{fig:sog.dendro_heat}.
Ward's method aims to minimize the within cluster variance at each
level of the hierarchy. At each step, the pairing which results in the
minimum increase in the variance is chosen. These increases are
measured as a weighted squared distance between cluster centers. 
We chose Ward's method over other linkage techniques
based on its tendency to produce clusters of comparable size at each
level of the hierarchy. 

At the most coarse resolution of two categories, we see 
in Fig.~\ref{fig:sog.dendro_heat} that 
one sided contests are distinguished from games that remain closer,
and repeated analysis using $k$-means clustering suggests the
same presence of two major clusters.

As we are interested in creating a taxonomy of more
particular, interpretable game shapes,
we opt to make cuts as $\tavg{\rho(d)}$ first falls below
an integer number of points, as 
shown in Fig.~\ref{fig:sog.point_breaks}
(we acknowledge that $\tavg{\rho(d)}$ does not perfectly decrease monotonically).
As indicated by the red vertical lines,
average intra-cluster point differences of 12, 11, 10, 9, and 8
correspond to 9, 18, 30, 71, and 157 distinct clusters.
Our choice, which is tied to a natural game score,
has a useful outcome of making the number of clusters approximately double
with every single point in average score differential.

\subsection{Taxonomy of 18 motifs for real AFL games}
\label{subsec:18motifs}

\begin{figure}[tbp!]
  \centering 
  \includegraphics[width=\columnwidth]{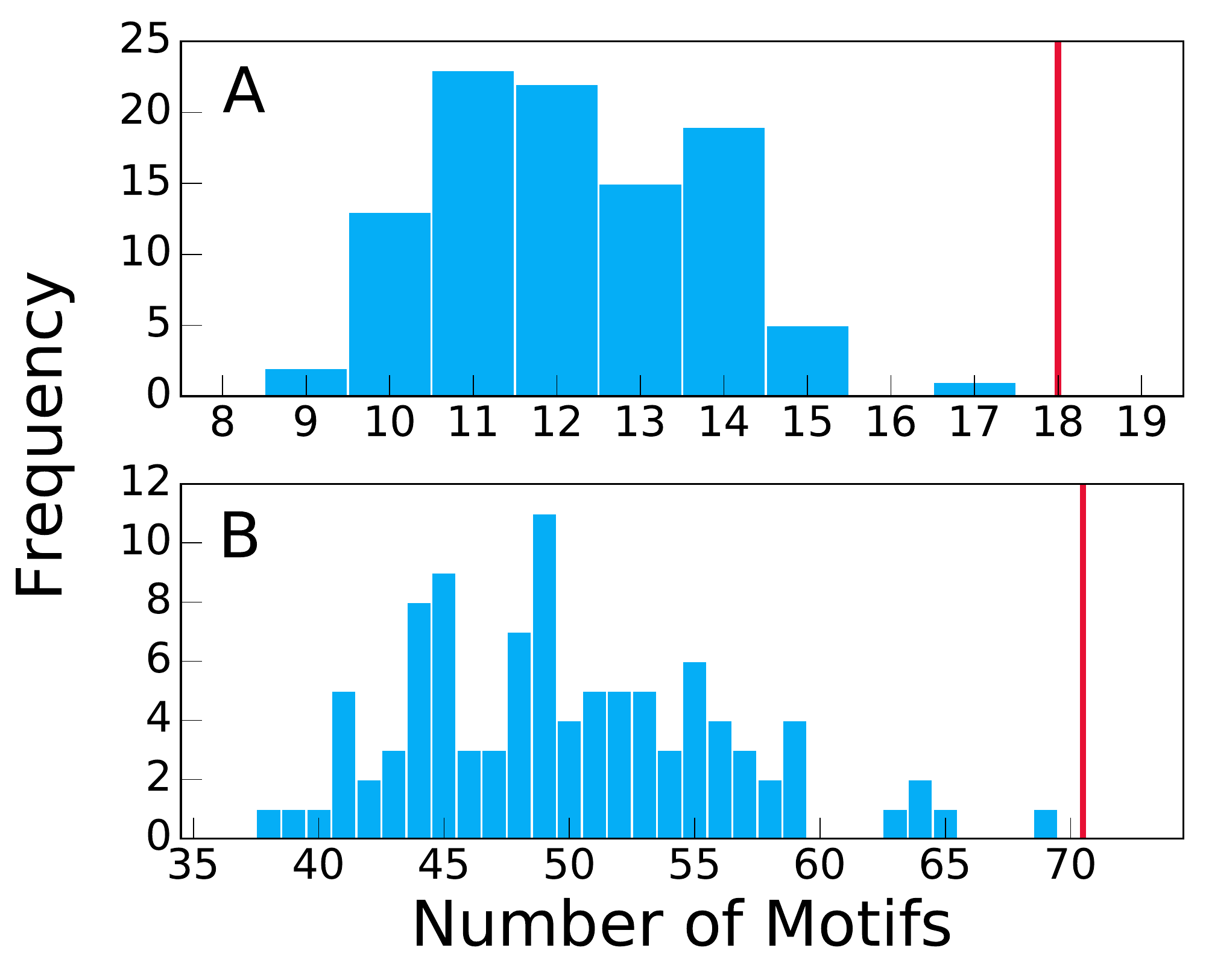}
  \caption{
    Histograms of the number of motifs produced by 100 ensembles
    of 1,310 games
    using the random walk null model, and evaluating at 
    11 
    and 
    9 point cutoffs  (\textbf{A} and \textbf{B})
    as described in Sec.~\ref{subsec:sog.coarsegraining}.
    For real games, we obtain by comparison 18 and 71 motifs (vertical
    red lines in \textbf{A} and \textbf{B}),
    which exceeds all 100 motif numbers in both cases
    and indicates AFL game stories are more diverse
    than our null model would suggest.
  }
  \label{fig:sog.randomwalk_motif_numbers}
\end{figure}

\begin{figure}[tbp!]
  \centering
  \includegraphics[width=\columnwidth]{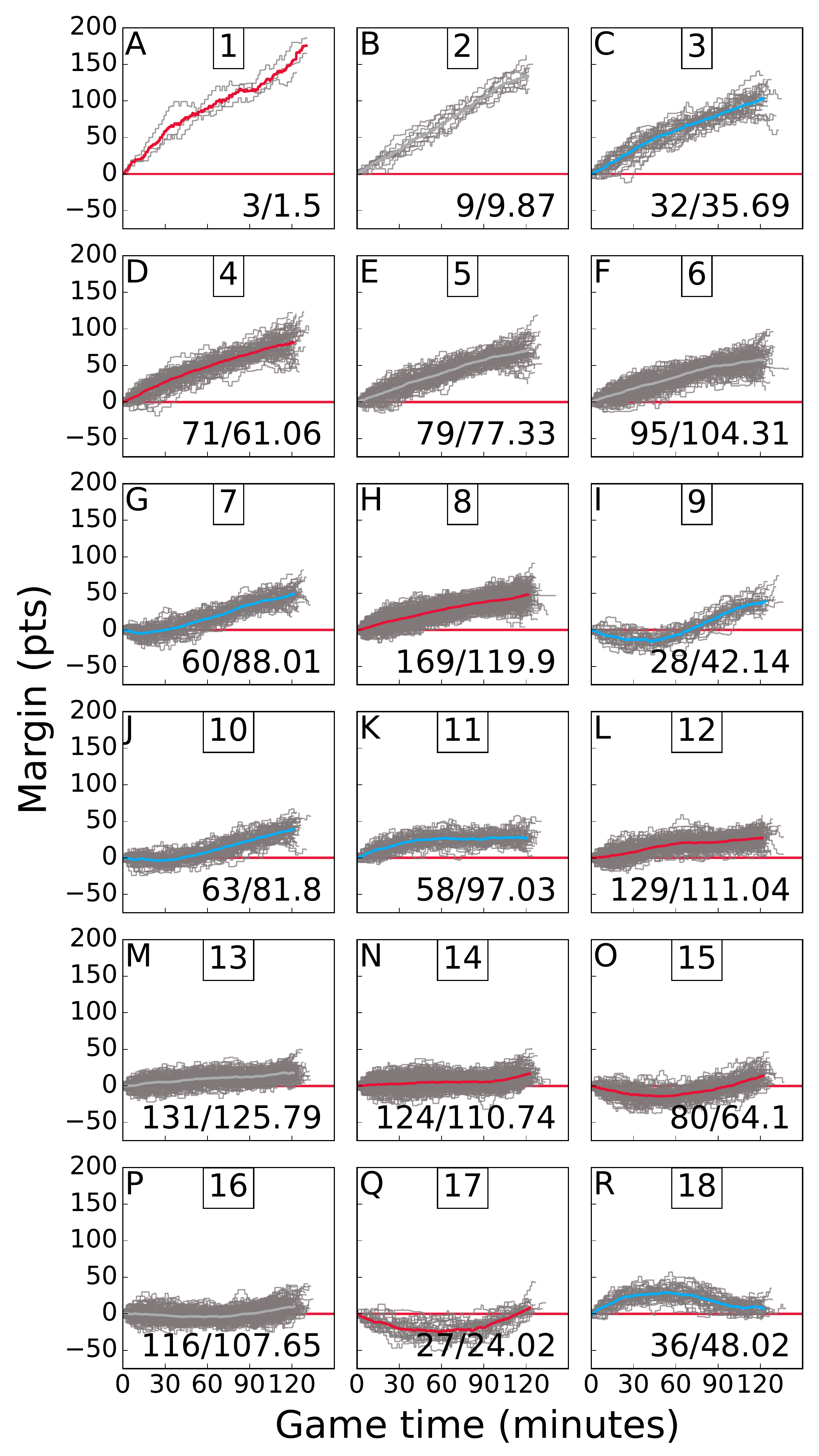}
        \caption{ 
    Eighteen game motifs as determined by
    performing hierarchical clustering analysis
    and finding when
    the average intra-cluster
    game distance $\tavg{\rho}$
    first drops below 11 points.
    In each panel, the main curves are the motifs---the average of
    all game stories (shown as light gray curves in background) within each cluster,
    and we arrange clusters in order of the motif winning margin.
    All motifs are shown with the same axis limits.
    Numbers of games within each cluster
    are indicated in the bottom right corner of each panel
    along with the average number of the nearest biased random walk games
    (normalized per 1,310).
    Motif colors correspond to relative abundance of
    real versus random game ratio $\gameratio$ as
    red: $\gameratio \ge 1.1$;
    gray: $0.9 < \gameratio < 1.1$;
    and 
    blue: $\gameratio \le 0.9$.
    See Fig.~\ref{fig:sog.realmotifs18ratio-random}
    for the same motifs reordered by real game to random ratio.
  }
  \label{fig:sog.biased18motifs}
\end{figure}

In the remainder of section~\ref{sec:sog.gamemotifs},
we show and explore in some depth the taxonomies provided by 18 and 71 motifs at the 11 and 9
point cutoff scales.

We first show that for both cutoffs, the number of motifs produced 
by the biased random walk null model
is typically well below the number observed for the real game.
In Fig.~\ref{fig:sog.randomwalk_motif_numbers}, we show
histograms of the number of motifs found in the 100 ensembles
of 1,310 null model games with the real game motif numbers of 18 and
71 marked by vertical red lines.
The number of random walk motifs is variable with both distributions
exhibiting reasonable spread, and
also in both cases, the maximum number of motifs is below the real game's
number of motifs.
These observations strongly suggest that AFL generates 
a more diverse set of game story shapes than
our random walk null model.

We now consider the 18 motif characterization
which we display in Fig.~\ref{fig:sog.biased18motifs} by plotting
all individual game stories in each cluster (light gray curves) 
and overlaying the average motif game story (blue/gray/red curves,
explained below).

All game stories are oriented so that the winning team aligns
with the positive vertical axis, i.e., $g_i(T) \ge 0$ 
(in the rare case of a tie, we orient the game story randomly).
and motifs are ordered by their final margin (descending).
In all presentations of motifs that follow,
we standardize final margin as the principle index of ordering.
We display the final margin index in the top center of each motif panel
to ease comparisons when motifs are ordered in other ways (e.g., by prevalence in the
null model).
We can now also connect back to the heat map of Fig.~\ref{fig:sog.dendro_heat}
where we use the same indices to mark the 18 motifs.

In the bottom right corner of each motif panel, we 
record two counts:
(1) the number of real games belonging to the motif's cluster;
and
(2) the average number of our ensemble of 100 $\times$ 1,310 biased random walk games 
(see Sec.~\ref{sec:sog.randomwalks})
which are closest to the motif according to~\Req{eq:sog.gamedist}.
For each motif,
we compute the ratio of real to random adjacent game stories, $\gameratio$,
and, as a guide, we color the motifs 
as 
\begin{itemize}
\item 
red if $\gameratio \ge 1.1$ (real game stories are more abundant);
\item 
gray if $0.9 < \gameratio < 1.1$ (counts of real and random game stories are
close);
and
\item 
blue if $\gameratio \le 0.9$ (random game stories are more abundant).
\end{itemize}

\begin{figure}[tbp!]
  \centering
  \includegraphics[width=\columnwidth]{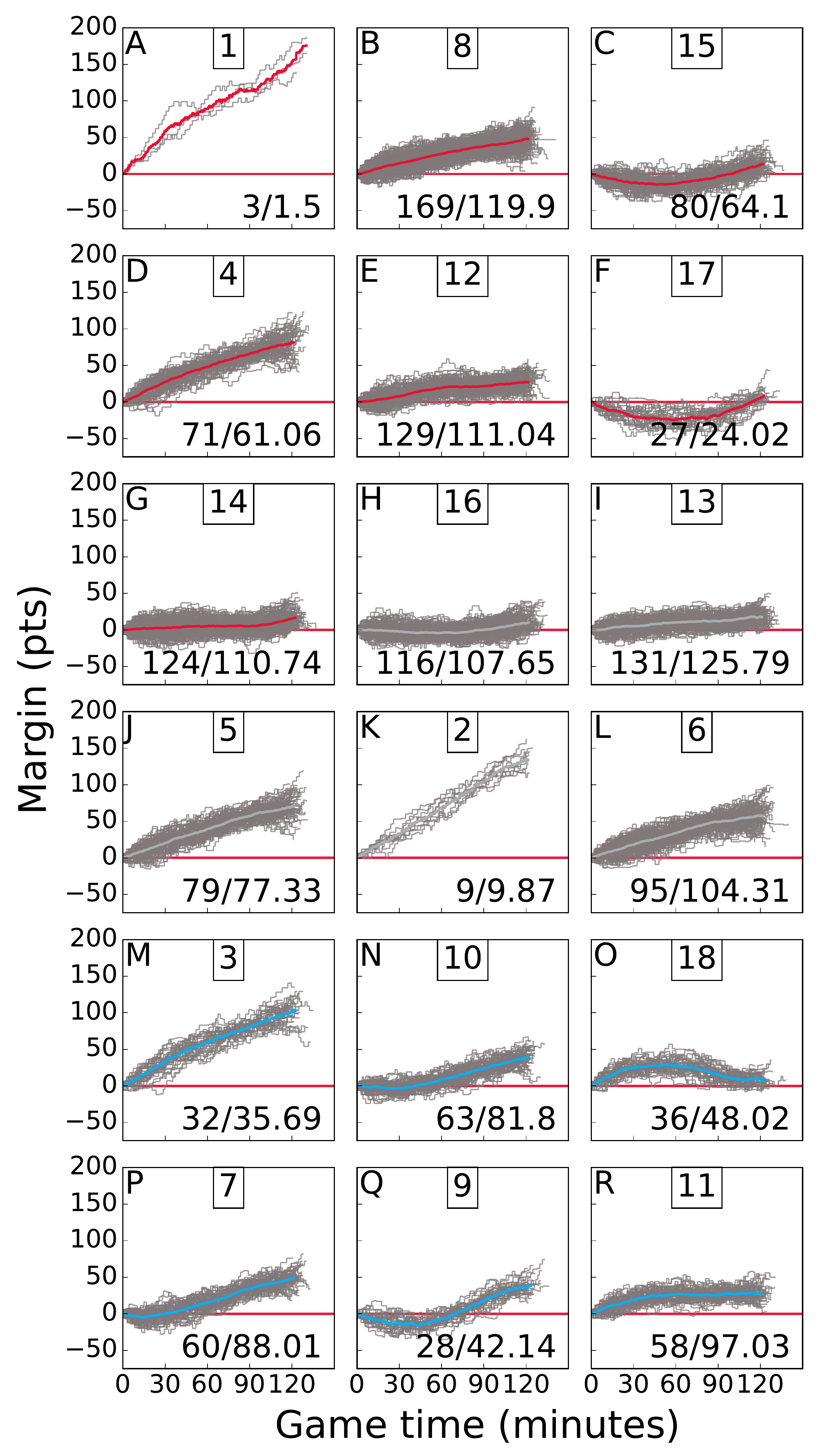}
  \caption{ 
    Real game motifs for an 11 point cut off as per
    Fig.~\ref{fig:sog.biased18motifs} 
    but
    reordered according to
    decreasing ratio of adjacent real to biased random games, $\gameratio$,
    and with 
    closest biased random walk rather than real game stories plotted
    underneath in light gray.
    See the caption for Fig.~\ref{fig:sog.biased18motifs} 
    for more details.
  }
  \label{fig:sog.realmotifs18ratio-random}
\end{figure}

We immediately observe that the number of games falling 
within each cluster is highly variable, 
with only 3 in the most extreme blowout motif (\#1, Fig.~\ref{fig:sog.biased18motifs}A/Fig.~\ref{fig:sog.realmotifs18ratio-random}A) 
and 169 in a gradual-pulling-away motif (\#8,
Fig.~\ref{fig:sog.biased18motifs}H/Fig.~\ref{fig:sog.realmotifs18ratio-random}B).

The average motif game stories in Fig.~\ref{fig:sog.biased18motifs}
provide us with the essence of each cluster,
and, though they do not represent any one real game,
they are helpful for the eye in distinguishing clusters.
Naturally, by applying further coarse-graining
as we do below, we 
will uncover a richer array of more specialized motifs.

Looking at Figs.~\ref{fig:sog.biased18motifs}
and~\ref{fig:sog.realmotifs18ratio-random},
we now clearly see a continuum of game shapes ranging from
extreme blowouts (motif \#1) to extreme comebacks,
both successful (motif \#17) and failed (motif \#18).
We observe that while some motifs have qualitatively similar story lines, 
a game motif that has a monotonically increasing score
differential that ends with a margin of 200 (\#1) is certainly different
from one with a final margin of 50 (\#6).

In considering this induced taxonomy of 18 game motifs,
we may interpret the following groupings:
\begin{itemize}
\item 
  \#1--\#6, \#8: One-sided, runaway matches;
\item 
  \#9: Losing early on, coming back, and then pulling away;
\item 
  \#7 and \#10: Initially even contests with one side eventually breaking away;
\item 
  \#11 and \#12: One team taking an early lead and then holding on for the rest of the
  game;
\item 
  \#13, \#14, and \#16: Variations on tight contests;
\item 
  \#15 and \#17: Successful comebacks;
\item 
  \#18: Failed comebacks.
\end{itemize}
We note that the game stories attached to each motif might
not fit these descriptions---we are only categorizing motifs.
As we move to finer grain taxonomies, the neighborhood around
motifs diminishes and the connection between the shapes of
motifs will become increasingly congruent with its constituent games.

The extreme blowout motif for real games
has relatively fewer adjacent random walk game stories
(Fig.~\ref{fig:sog.realmotifs18ratio-random}A),
as do
the two successful comeback motifs
(Fig.~\ref{fig:sog.realmotifs18ratio-random}C
and Fig.~\ref{fig:sog.realmotifs18ratio-random}F),
and games with a lead developed by half time
that then remains stable
(Fig.~\ref{fig:sog.realmotifs18ratio-random}E).
A total of 5 motifs show 
a relatively even balance between real
and random (i.e., within 10\%)
including two of the six motifs with the tightest
finishes
(Figs.~\ref{fig:sog.realmotifs18ratio-random}H
and \ref{fig:sog.realmotifs18ratio-random}I).
Biased random walks most overproduce 
games in which an early loss is turned around strongly
(Fig.~\ref{fig:sog.realmotifs18ratio-random}Q)
or an early lead is maintained
(Fig.~\ref{fig:sog.realmotifs18ratio-random}R).
In terms of game numbers behind motifs,
we find a reasonable balance with 
603 (46.0\%) having $\gameratio \ge 1.1$ (7 motifs),
430 (32.8\%) with $0.9 < \gameratio < 1.1$ (5 motifs),
and
277 (21.1\%) with $\gameratio \le 0.9$ (6 motifs).

Depending on the point of view of the fan 
and again at this level of 18 motifs,
we could argue that certain real AFL games that feature more often
that our null model would suggest are more or less `interesting'.
For example, we see some dominating wins are relatively more abundant in the real
game (\#1, \#2, and \#4).
While such games are presumably gratifying for fans of the team handing out the
`pasting',
they are likely deflating for the supporters of the losing team.
And a neutral observer may or may not enjoy the spectacle of 
a superior team displaying their prowess.
Real games do exhibit relatively more of the two major comeback motifs
(\#15 and \#17)---certainly exciting in nature---though less of the failed comebacks (\#18).

\begin{figure*}[tbp!]
  \centering
  \includegraphics[width=0.92\textwidth]{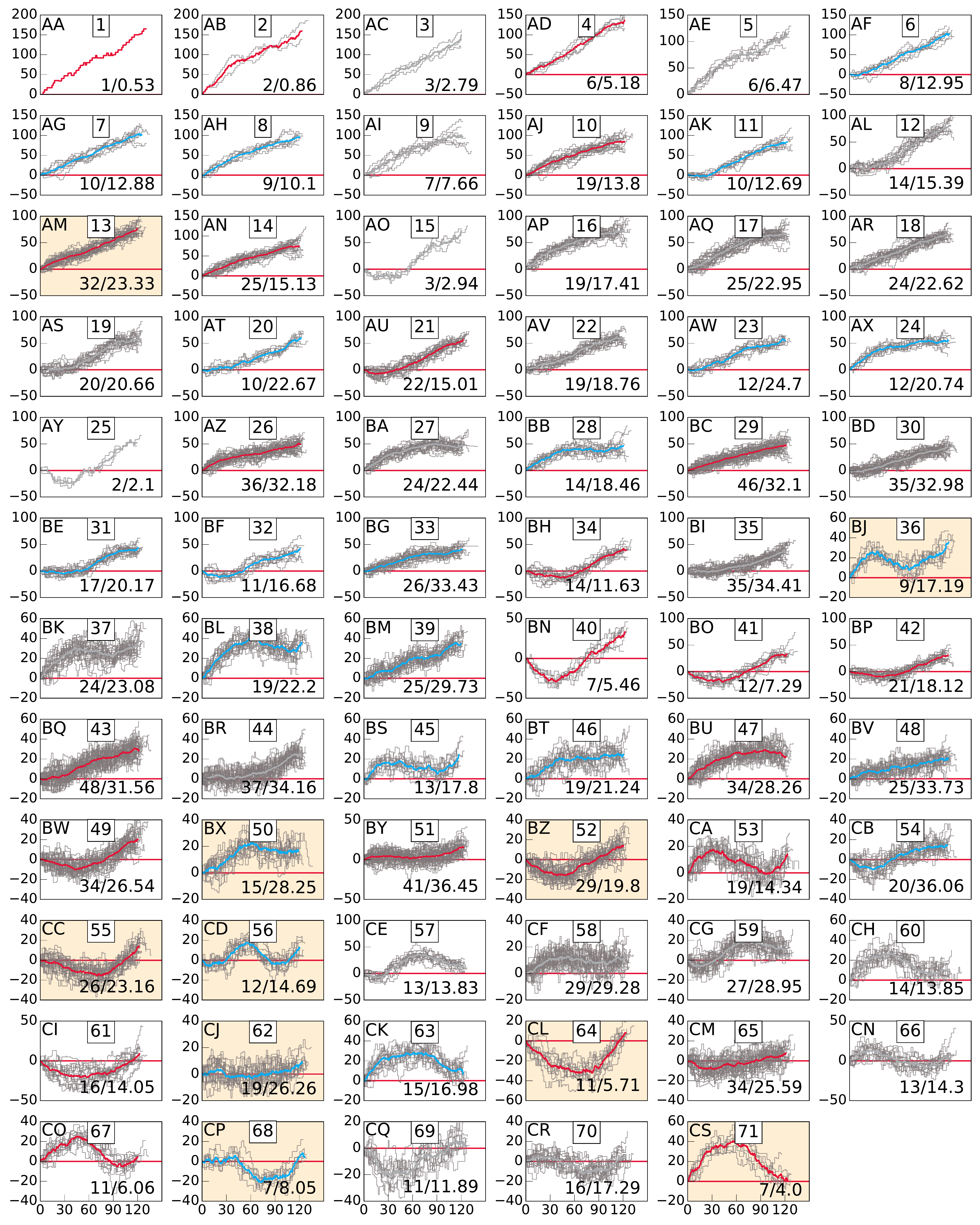}
    \caption{ 
    Seventy-one distinct game motifs as determined by
    hierarchical clustering analysis with a threshold of nine points,
    the fourth cutoff shown in 
    Fig.~\ref{fig:sog.point_breaks} and 
    described in 
    Sec.~\ref{sec:sog.gamemotifs}.
    Motifs are ordered by their final margin, highest to lowest,
    and real game stories are shown in the background of each motif.
    Cutoffs for motif colors red, gray, and blue correspond to
    real-to-random ratios 1.1 and 0.9, and the top number indicates
    motif rank according to final margin.
    The same process applied to the biased random walk model for our
    100 simulations
    typically yields only 45 to 50 motifs (see Fig.~\ref{fig:sog.randomwalk_motif_numbers}).
    We discuss the ten highlighted motifs in the main text
    and note that we have allowed
    the vertical axis limits to vary.
  }
  \label{fig:sog.biased71motifs-final-margin-order}
\end{figure*}

\begin{figure*}[tbp!]
  \centering
  \includegraphics[width=0.92\textwidth]{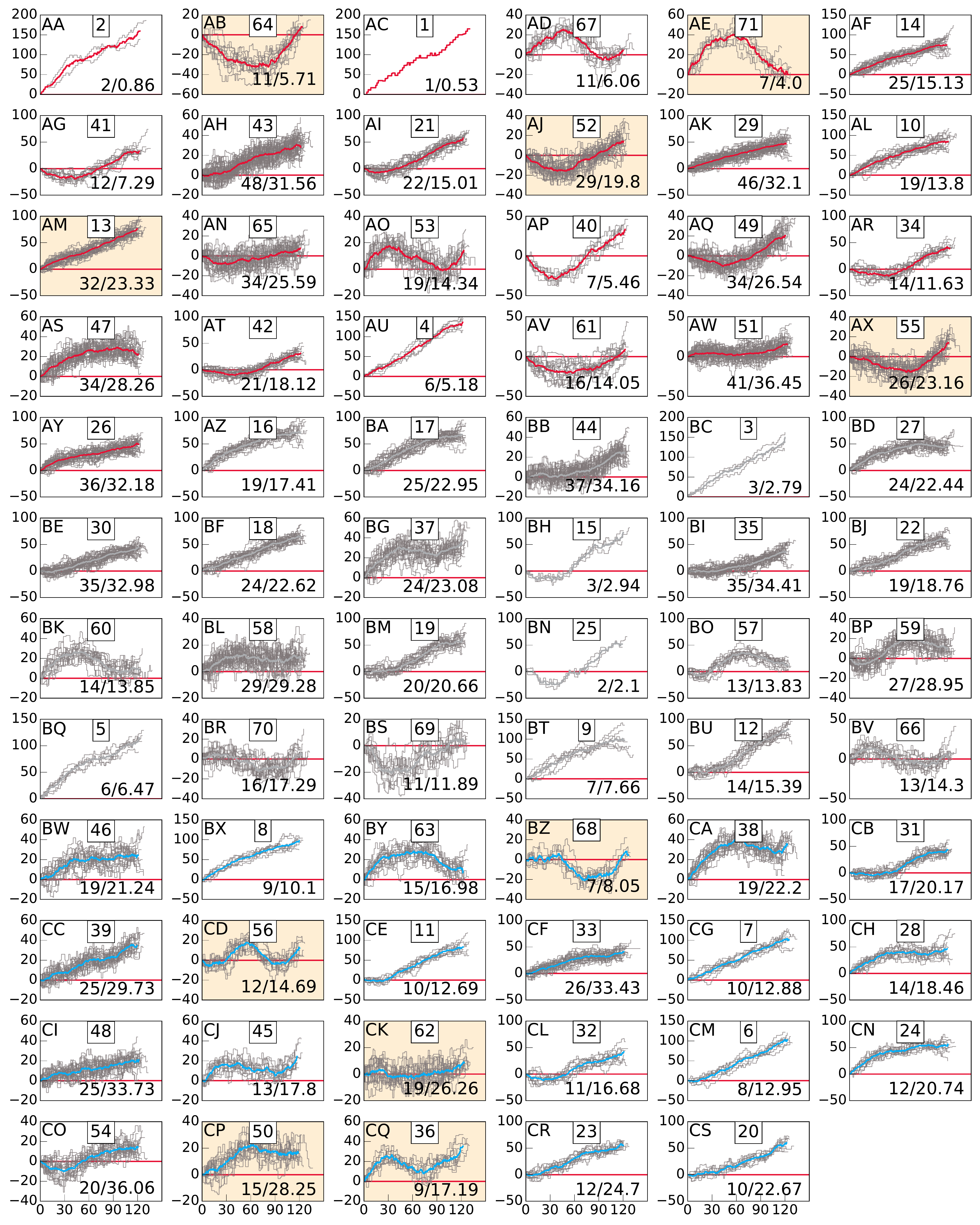}
    \caption{ 
    Motifs (red curves) from
    Fig.~\ref{fig:sog.biased71motifs-final-margin-order}
    rearranged in order of descending ratio of the number of real games
    to the number of adjacent biased random walk games,
    as described in Sec.~\ref{subsec:18motifs},
    and adjacent real game stories are shown in gray for each motif.
  }
  \label{fig:sog.biased71motifs-ratio-order}
\end{figure*}

\subsection{Taxonomy of 71 motifs for real AFL games}
\label{subsec:71motifs}

Increasing our level of resolution corresponding to 
an average intra-cluster game distance of $\tavg{\rho} = 9$, 
we now resolve the AFL game story ecology into 71 clusters.
We present all 71 motifs in
Figs.~\ref{fig:sog.biased71motifs-final-margin-order} and
\ref{fig:sog.biased71motifs-ratio-order},
ordering by final margin and real-to-random game story ratio
$\gameratio$ respectively (we will refer to motif number and
Fig.~\ref{fig:sog.biased71motifs-ratio-order}
so readers may easily connect to the orderings in both figures).
With a greater number of categories, we naturally see a more even distribution
of game stories across motifs with a minimum of 1
(Motif \#1, Fig.~\ref{fig:sog.biased71motifs-ratio-order}AC)
and
a maximum of 
48
(Motif \#43, Fig.~\ref{fig:sog.biased71motifs-ratio-order}AH).

As for the coarser 18 motif taxonomy, we again observe a 
mismatch between real and biased random walk games.
For example, motif \#14 (Fig.~\ref{fig:sog.biased71motifs-ratio-order}AF) 
is an average
of 25 real game stories compared with on average 15.13 adjacent biased
random walks
while  motif \#20 (Fig.~\ref{fig:sog.biased71motifs-ratio-order}CS) 
has $\gameratio$=10/22.67.
Using our 10\% criterion, we see 25 motifs have $\gameratio \ge 1.1$ 
(representing 553 games or 42.2\%),
23 have $0.9 < \gameratio < 1.1$ (420 games, 32.0\%),
and
the remaining 
23 have $\gameratio \le 0.9$ (337 games, 25.7\%).
Generally, we again see blowouts are more likely
in real games.
However, we also find some kinds of comeback motifs are also more prevalent ($\gameratio \ge 1.1$)
though not strongly in absolute numbers; these include
the failed comebacks in motifs
\#67 (Fig.~\ref{fig:sog.biased71motifs-ratio-order}AD)
and
\#71 (Fig.~\ref{fig:sog.biased71motifs-ratio-order}AE),
and
the major comeback in motif
\#64 (Fig.~\ref{fig:sog.biased71motifs-ratio-order}AB).

\begin{figure}[tbp!]
  \centering
  \includegraphics[width=\columnwidth]{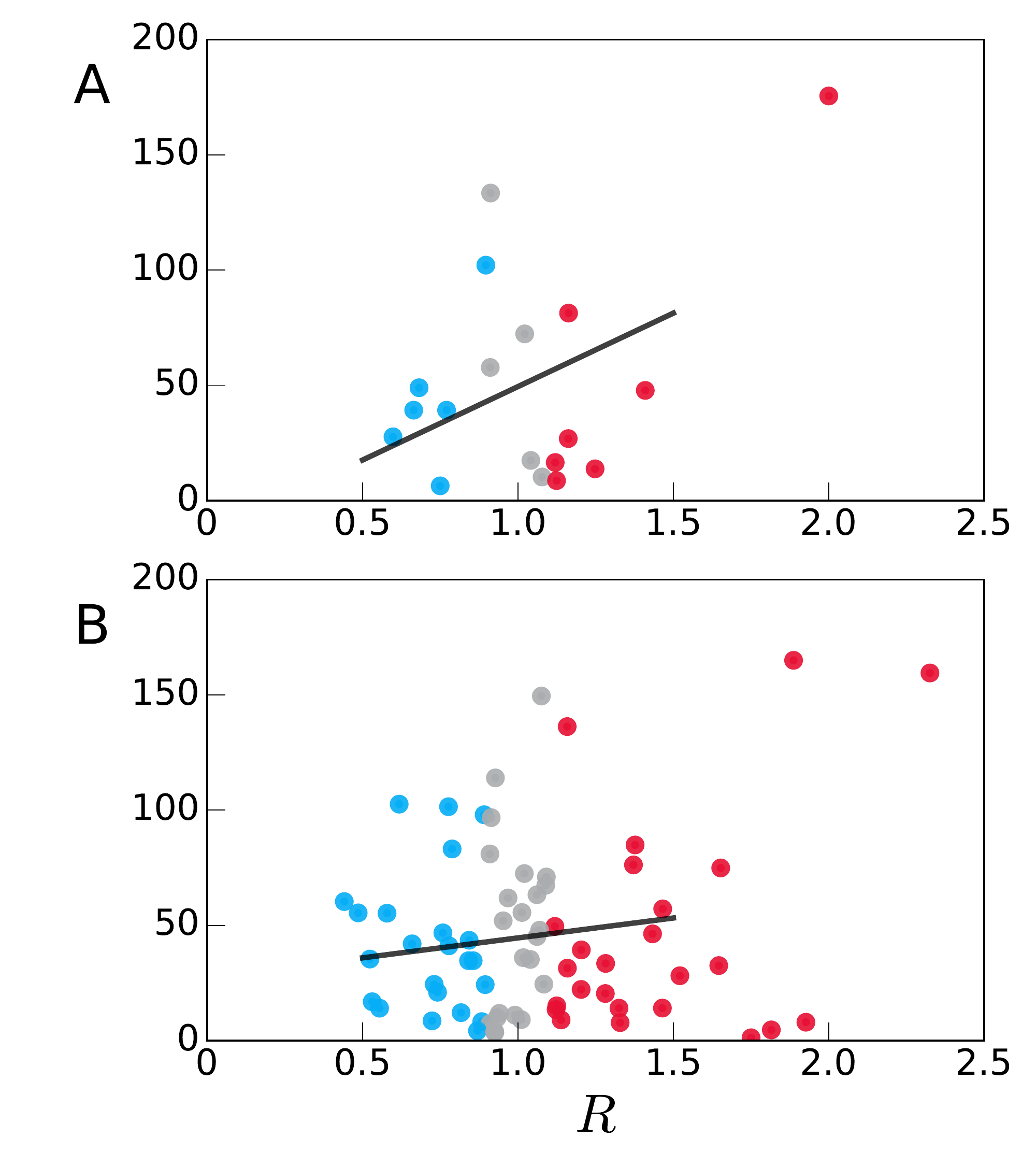}
  \caption{ 
    Final margin of motifs as a function of real-to-random ratio
    $\gameratio$
    for real AFL games
    at the 
    18 
    and 71 
    motif levels,
    panels
    \textbf{A}
    and 
    \textbf{B}
    respectively, with linear fits.
    On the right of each plot, extreme blowout motifs ending in high margins have
    no or relatively few adjacent random walks.
    (red, $\gameratio \ge 1.1$).
    On the left, game stories are more well represented by random walks
    (blue, $\gameratio \le 0.9$).
    There is considerable variation however, particularly in the 71
    motif case, and
    we certainly see some close finishes with $\gameratio \ge 1$
    (e.g., the massive comeback, 
    motif \#71,
    Fig.~\ref{fig:sog.biased71motifs-ratio-order}AE).
  }
  \label{fig:sog.biasedRatio_vs_Margin_18_and_71}
\end{figure}

In Fig.~\ref{fig:sog.biasedRatio_vs_Margin_18_and_71}, 
we  give summary plots for the 18 and 71 motif taxonomies
with motif final margin as a function of the 
of the real-to-random ratio $\gameratio$.
The larger final margins of the blowout games feature on the right of
these plots ($\gameratio \ge 1.1$), and, in moving to the left,
we see a gradual tightening of games
as shapes become more favorably produced by the random null model 
($\gameratio \le 0.9$).
The continuum of game stories is also reflected
in the basic similarity of the two plots
in Fig.~\ref{fig:sog.biasedRatio_vs_Margin_18_and_71}, 
made as they are for
two different levels of coarse-graining.

Returning to Figs.~\ref{fig:sog.biased71motifs-final-margin-order}
and~\ref{fig:sog.biased71motifs-ratio-order}, we highlight ten
examples in both reinforcements and refinements of motifs seen at the 18 motif level.
We frame them as follows (in order of decreasing $\gameratio$ and
referencing Fig.~\ref{fig:sog.biased71motifs-ratio-order}):
\begin{itemize}
\item 
  Fig.~\ref{fig:sog.biased71motifs-ratio-order}AB,
  \#64 ($\gameratio=11/5.71$): 
  The late, great comeback;
\item
  Fig.~\ref{fig:sog.biased71motifs-ratio-order}AE,
  \#71 ($\gameratio=7/4.00$): 
  The massive comeback that just falls short;
\item 
  Fig.~\ref{fig:sog.biased71motifs-ratio-order}AJ,
  \#52 ($\gameratio=29/19.80$): 
   comeback over the first half connecting into a blowout in the
   second
   (the winning team may be said to have `Turned the corner');
\item 
  Fig.~\ref{fig:sog.biased71motifs-ratio-order}AM,
  Motif \#13 ($\gameratio=32/23.33$): 
  an exemplar blowout (and variously a shellacking, thrashing, or hiding);
\item 
  Fig.~\ref{fig:sog.biased71motifs-ratio-order}AX,
  \#55 ($\gameratio=26/23.16$): 
  Rope-a-dope (taking steady losses and then surging late);
\item 
  Fig.~\ref{fig:sog.biased71motifs-ratio-order}BZ,
  \#68 ($\gameratio=7/8.05$): 
  Hold-slide-hold-surge;
\item 
  Fig.~\ref{fig:sog.biased71motifs-ratio-order}CD,
  \#56 ($\gameratio=12/14.69$): 
  See-saw battle;
\item 
  Fig.~\ref{fig:sog.biased71motifs-ratio-order}CK,
  \#62 ($\gameratio=19/26.26$): 
  The tightly fought nail-biter (or heart stopper);
\item 
  Fig.~\ref{fig:sog.biased71motifs-ratio-order}CP,
  \#50 ($\gameratio=15/28.25$): 
  Burn-and-hold
  (or the game-manager, or the always dangerous playing not-to-lose);
\item 
  Fig.~\ref{fig:sog.biased71motifs-ratio-order}CQ,
  \#36 ($\gameratio=9/17.19$): 
  Surge-slide-surge.
\end{itemize}
These motifs may also be grouped according to
the number of `acts' in the game.
Motif \#53 (Fig.~\ref{fig:sog.biased71motifs-ratio-order}AO),
for example, is a three-act story
while motifs 
\#56 (Fig.~\ref{fig:sog.biased71motifs-ratio-order}CD)
and 
\#68 (Fig.~\ref{fig:sog.biased71motifs-ratio-order}BZ)
exhibit four acts.
We invite the reader to explore the rest of the motifs
in 
Fig.~\ref{fig:sog.biased71motifs-ratio-order}.

\section{Predicting games using shapes of stories}
\label{sec:sog.prediction}

Can we improve our ability to predict the outcome of a game in
progress by knowing how games with similar stories played out in the
past? Does the full history of a game help us gain any predictive
power over much simpler game state descriptions such as the current
time and score differential? 
In this last section, we explore
prediction as informed by game stories, a natural application. 

Suppose we are in the midst of viewing a new game.
We know the game story
$g_{\textnormal{obs}}$ from the start of the game
until  the current game time $t < T_{\textnormal{obs}}$,
where
$T_{\textnormal{obs}}$
is the eventual length of game
(and is another variable which we could potentially predict).
In part to help with presentation and analysis, we will use minute
resolution (meaning $t=60n$ for $n = 0, 1, 2, \ldots$).
Our goal is to use our database of
completed games---for which of course we know the eventual outcomes---to
predict the final margin of our new game, $g_{\textnormal{obs}}(T_{\textnormal{obs}})$.

We create a prediction model with two parameters:
(1)
$N$: the desired number of analog games closest to our present game;
and
(2)
$M$: the number of minutes going back from the current time
for which we will measure the distance between games.
For a predictor, we simply average the final margins of the $N$ closest analog games
to $g_{\textnormal{obs}}$ over the interval $[t-60M,t]$.
That is, at time $t$, we predict the
final margin of $g_{\textnormal{obs}}$, $F$,
using $M$ minutes of memory and $N$ analog games as:
\begin{equation}
  F(g_{\textnormal{obs}}, t/60, M, N) 
  = 
  \frac{1}{N}
  \sum_{i \in \Omega(g_{\textnormal{obs}},t/60,M,N)}
  g_{i} (T_{i}),
  \label{eq:sog.prediction}
\end{equation}
where 
$\Omega(g_{\textnormal{obs}},t/60,M,N)$
is the set of indices for the $N$ games closest to the current game
over the time span $[t-60M,t]$,
and 
$T_i$ 
is the final
second of game 
$i$.

\begin{figure}[tbp!]
  \centering 
  \includegraphics[width=\columnwidth]{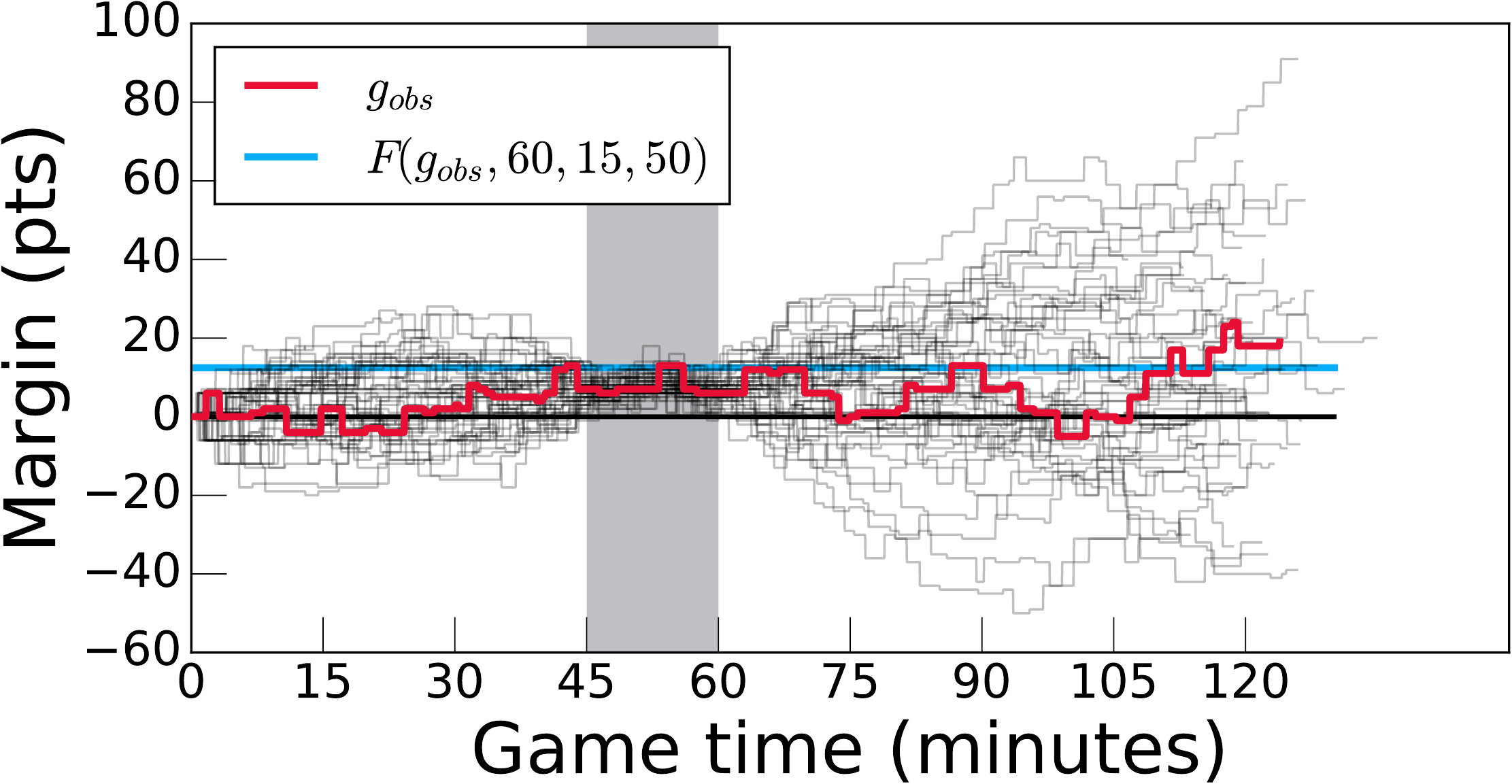}
  \caption{
    Illustration of our prediction method 
    given in \Req{eq:sog.prediction}.
    We start with a game story $g_{\textnormal{obs}}$ (red curve)
    for which we know up until, for this example, 60 minutes ($t=3600$).
    We find the $N=50$ closest game stories based on matching
    over the time period 45 to 60 minutes (memory $M=15$),
    and show these as gray curves.
    We indicate the average final score
    $F(g_{\textnormal{obs}}, t/60, M, N)$
    for these analog games 
    with the horizontal blue curve.
  }
  \label{fig:sog.prediction_example}
\end{figure}

\begin{figure*}[tbp!]
  \centering 
  \includegraphics[width=1.1\textwidth]{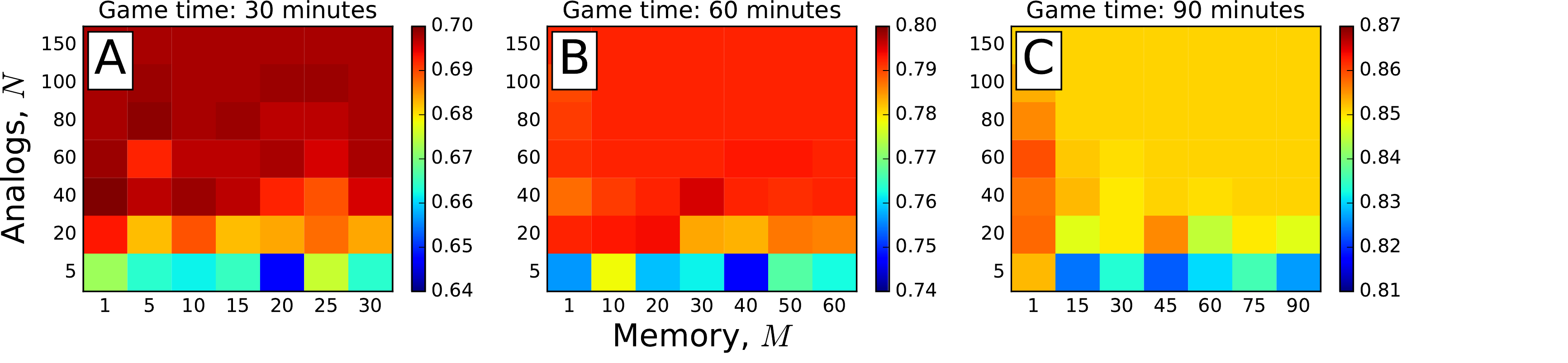}
  \caption{
    Fraction of games correctly predicted using the average final
    score of $N$ analog games, with adjacency evaluated over the last
    $M$ minutes 
    at the three game times of 
    \textbf{A.} 30,
    \textbf{B.} 60, 
    and 
    \textbf{C.} 90 minutes.
    Increasing the number of analogs provides the strongest benefit
    for prediction
    while increasing memory either degrades or does not improve
    performance.
    Because prediction improves as a game is played out,
    the color bars cover the same span of accuracy (0.06) but
    with range translated appropriately.
  }
  \label{fig:sog.memory_vs_analogs}
\end{figure*}

\begin{figure}[tbp!]  
  \centering
  \includegraphics[width=\columnwidth]{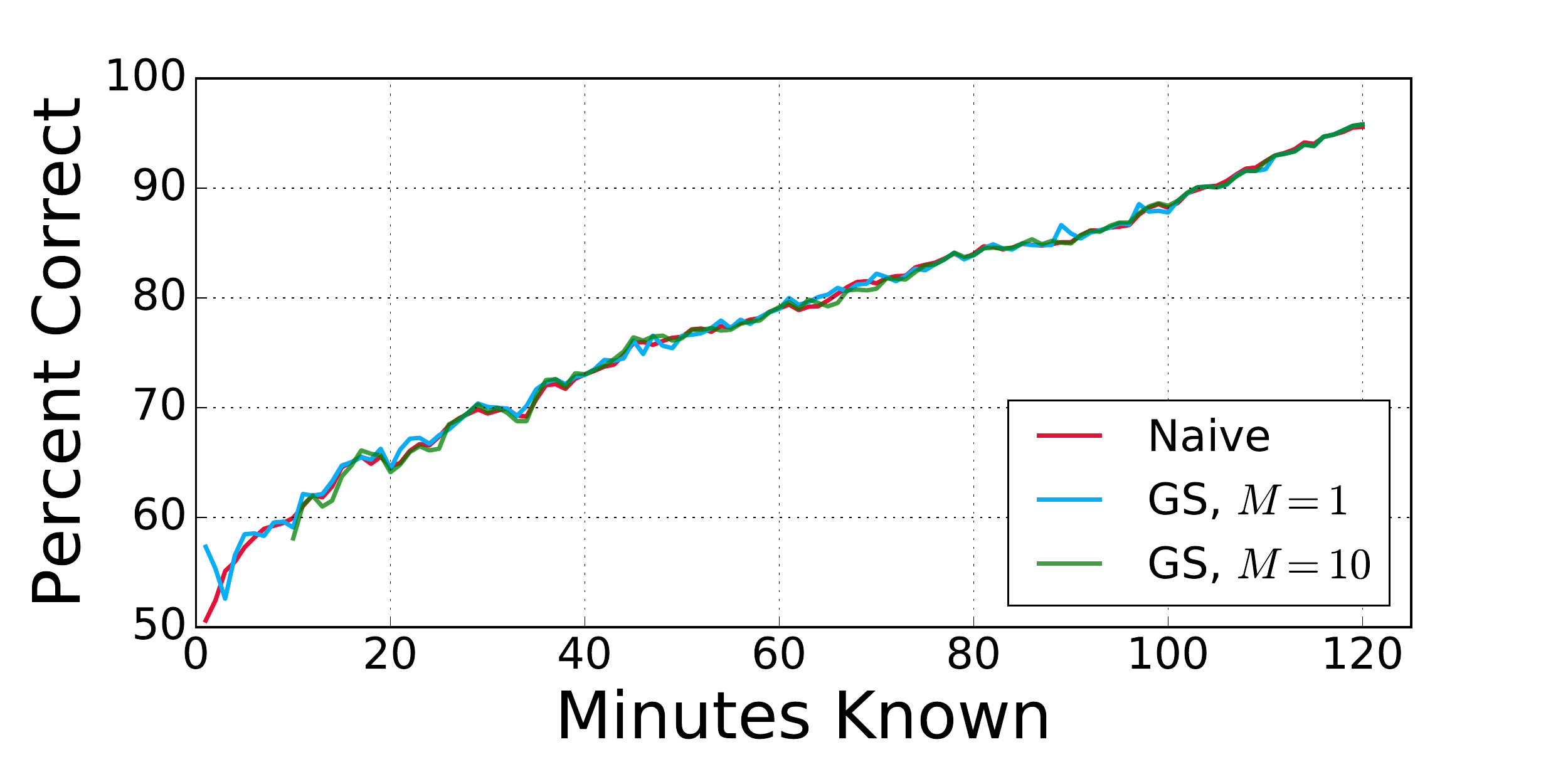}
  \caption{ 
    Prediction accuracy using the described game shape
    comparison model using $N=50$ analogs and a memories of $M=1$
    (blue curve) and $M=10$ (green curve), compared with the
    naive model of assuming that the current leader will ultimately
    win (red curve). 
  }
  \label{fig:sog.prediction_performance}
\end{figure}

For an example demonstration, in Fig.~\ref{fig:sog.prediction_example}, 
we attempt to predict the outcome of an example game story
given knowledge of its first 60 minutes (red curve)
and by finding the average final margin of the $N = 50$ closest
games over the interval 45 to 60 minutes ($M=15$, shaded gray region).
Most broadly, we see that our predictor here would correctly call
the winning team. 
At a more detailed level, the average final margin of the analog
games slightly underestimates the final margin
of the game of interest, and the range of outcomes
for the 50 analog games is broad with the final margin
spanning from around -40 to 90 points.

Having defined our prediction method, we now systematically test its
performance 
after 30, 60, and 90 minutes have elapsed in a game currently under way.
In aiming to find the best combination of memory and analog number, 
$M$ and $N$, 
we use~\Req{eq:sog.prediction} 
to predict the eventual winner of all 1,310 AFL games in our data set
at these time points.
First, as should be expected, the further a game has progressed,
the better our prediction.
More interestingly, in Fig.~\ref{fig:sog.memory_vs_analogs}
we see that for all three time points,
increasing $N$ elevates the prediction accuracy, 
while increasing $M$ has little and sometimes the opposite effect, 
especially for small $N$.
The current score differential serves as a stronger indicator of
the final outcome than the whole game story shape unfolded so far.
The recent change in scores---momentum---is also informative, 
but to a far lesser extent than the simple
difference in scores at time $t$.

Based on Fig.~\ref{fig:sog.memory_vs_analogs},
we proceed with $N=50$ analogs and two examples
of low memory: $M=1$ and $M=10$.
We compare with the naive model that, at any time $t$,
predicts the winner as being the current leader.

We see in Fig.~\ref{fig:sog.prediction_performance}
that there is essentially no difference in prediction
performance between the two methods.
Thus, memory does not appear to play a necessary role
in prediction for AFL games.  Of interest going forward
will be the extent to which other sports show the
same behavior.
For predicting the final score, we also observe that 
simple linear extrapolation performs 
well on the entire set of the AFL games (not shown).

Nevertheless, we have thus far found no compelling evidence for using game
stories in prediction, nuanced analyses incorporating game stories for
AFL and other professional sports may nevertheless yield substantive
improvements over these simple predictive models~\cite{peel2015a}.

\section{Concluding remarks}
\label{sec:sog.conclusion}

Overall, we find that the sport of Australian Rules Football presents
a continuum of game types ranging from dominant blowouts 
to last minute, major comebacks.
Consequently, and rather than uncovering an optimal number of game motifs, 
we instead apply coarse-graining to find a varying number of motifs depending on the degree
of resolution desired.

We further find that
(1) A biased random walk affords a reasonable null model for AFL game
stories;
(2) The scoring bias distribution may be numerically determined so
that the null model produces a distribution of final margins 
which suitably matches that of real games;
(3) Blowout and major comeback  motifs are much more strongly
represented in the real game whereas tighter games 
are generally (but not entirely) more favorably produced by a random model;
and
(4) AFL game motifs are overall more diverse than those of the random version.

Our analysis of an entire sport through its game story ecology 
could naturally be applied to other major sports such as American
football, Association football (soccer), basketball, and baseball.
A cross-sport comparison
for any of the above analysis would likely be interesting and
informative.
And at a macro scale, we could also 
explore the shapes of  win-loss progressions of 
franchises over years~\cite{morris2014a}.

It is important to reinforce that a priori, we were unclear as to
whether there would be distinct
clusters of games or a single spectrum, and one might imagine rough theoretical
justifications for both.  
Our finding of a spectrum conditions our expectations for other sports,
and also provides a stringent, nuanced test for more advanced explanatory
mechanisms beyond biased random walks, although we are wary of the potential difficulty involved in
establishing a more sophisticated and still defensible mechanism.

Finally, a potentially valuable future project
would be an investigation of the aesthetic quality
of both individual games and motifs
as rated by fans and neutral individuals~\cite{bryant2000a}. 
Possible sources of data would be (1) social media posts tagged as being
relevant to a specific game, and (2) information on game-related betting.
Would true fans rather see a boring blowout with their team on top
than witness a close game~\cite{pawlowski2013a,gantz1995a}?
Is the final margin the main criterion for an interesting game?
To what extent do large momentum swings engage an audience?
Such a study could assist in the implementation of new rules and
policies within professional sports.

\acknowledgments
We thank Eric Clark, Thomas McAndrew, and James Bagrow
for helpful
discussions.
PSD was supported by NSF CAREER Grant No. 0846668,
and CMD and PSD were supported by NSF BIGDATA Grant No. 1447634.

\bibliographystyle{unsrtabbrv}

\begin{thebibliography}{10}

\bibitem{billings2013a}
A.~C. Billings, editor.
\newblock {\em Sports {Media}: {Transformation}, {Integration}, {Consumption}}.
\newblock Routledge, New York etc., reprint edition edition, Sept. 2013.

\bibitem{bryant1994a}
J.~Bryant, S.~C. Rockwell, and J.~W. Owens.
\newblock ``{B}uzzer {B}eaters'' and ``{B}arn {B}urners'': {T}he effects on
  enjoyment of watching the game go ``down to the wire''.
\newblock {\em Journal of Sport and Social Issues}, 18(4):326--339, Nov. 1994.

\bibitem{gantz1995a}
W.~Gantz and L.~A. Wenner.
\newblock Fanship and the television sports viewing experience.
\newblock {\em Sociology of Sport Journal}, 12:56--74, 1995.

\bibitem{hugenberg2008a}
L.~W. Hugenberg, P.~M. Haridakis, and A.~C. Earnheardt.
\newblock {\em Sports {Mania}: {Essays} on {Fandom} and the {Media} in the 21st
  {Century}}.
\newblock McFarland, Jefferson, N.C, July 2008.

\bibitem{elderton1945a}
W.~Elderton.
\newblock Cricket scores and some skew correlation distributions: ({A}n
  arithmetical study).
\newblock {\em Journal of the Royal Statistical Society}, 108:1--11, 1945.

\bibitem{wood1945a}
G.~H. Wood.
\newblock Cricket scores and geometrical progression.
\newblock {\em Journal of the Royal Statistical Society}, 108:12--40, 1945.

\bibitem{colwell1982a}
D.~J. Colwell and J.~R. Gillett.
\newblock The random nature of cricket and football results.
\newblock {\em The Mathematical Gazette}, 66:137--140, 1982.

\bibitem{bocskocsky2014a}
A.~Bocskocsky, J.~Ezekowitz, and C.~Stein.
\newblock The hot hand: {A} new approach to an old ``fallacy''.
\newblock {\em MIT Sloan Sports Analytics Conference}, March 2014.

\bibitem{ribeiro2012a}
H.~V. Ribeiro, S.~Mukherjee, and X.~H.~T. Zeng.
\newblock Anomalous diffusion and long-range correlations in the score
  evolution of the game of cricket.
\newblock {\em Phys. Rev. E}, 86:022102, 2012.

\bibitem{gabel2012a}
A.~Gabel and S.~Redner.
\newblock Random walk picture of basketball scoring.
\newblock {\em Journal of Quantitative Analysis in Sports}, 8:1--20, 2012.

\bibitem{clauset2015a}
A.~Clauset, M.~Kogan, and S.~Redner.
\newblock Safe leads and lead changes in competitive team sports.
\newblock {\em Phys. Rev. E}, 91:062815, 2015.

\bibitem{merritt2014a}
S.~Merritt and A.~Clauset.
\newblock Scoring dynamics across professional team sports: {T}empo, balance
  and predictability.
\newblock {\em EPJ Data Science}, 3, 2014.

\bibitem{jesaulenko1970a}
{A}{F}{L} --- {A}lex {J}esaulenko {M}ark of the {Y}ear 1970 {G}rand {F}inal.
\newblock Available online at
  \url{https://www.youtube.com/watch?v=uRr_HSjF6Hg}; accessed on June 24, 2015.

\bibitem{afl_tables_www.afltables.com_2014}
{A}{F}{L} {T}ables: {A}{F}{L}-{V}{F}{L} match, player and coaching stats,
  records and lists, 2014.
\newblock Obtained online at
  \href{http://www.afltables.com}{http://www.afltables.com}, accessed May 15,
  2015.

\bibitem{australian_football_league_laws_2015}
{Australian Football League}.
\newblock Laws of {A}ustralian {R}ules {F}ootball, 2015.
\newblock Obtained online at
  \href{http://www.afl.com.au/laws}{http://www.afl.com.au/laws}, accessed May
  15, 2015.

\bibitem{thompson2008a}
P.~Thompson.
\newblock Learning by doing.
\newblock Working {Paper} 0806, Florida International University, Department of
  Economics, 2008.

\bibitem{berger2011a}
J.~Berger and D.~Pope.
\newblock Can losing lead to winning?
\newblock {\em Management Science}, 57(5):817--827, Apr. 2011.

\bibitem{feller1971a}
W.~Feller.
\newblock {\em An Introduction to Probability Theory and Its Applications},
  volume~II.
\newblock John Wiley \& Sons, New York, second edition, 1971.

\bibitem{kiley2015a}
D.~P. Kiley, A.~J. Reagan, L.~Mitchell, C.~M. Danforth, and P.~S. Dodds.
\newblock The game story space of professional sports: {A}ustralian {R}ules
  {F}bootball.
\newblock Draft version of the present paper using pure random walk null model.
  Available online at
  \href{http://arxiv.org/abs/1507.03886v1}{http://arxiv.org/abs/1507.03886v1}.
  Accesssed January 17, 2016, 2015.

\bibitem{ward1963a}
J.~H. Ward.
\newblock Hierarchical grouping to optimize an objective function.
\newblock {\em Journal of the American Statistical Association}, 58:236--244,
  1963.

\bibitem{peel2015a}
L.~Peel and A.~Clauset.
\newblock Predicting sports scoring dynamics with restoration and
  anti-persistence.
\newblock Available online at
  \href{http://arxiv.org/abs/1504.05872}{http://arxiv.org/abs/1504.05872},
  2015.

\bibitem{morris2014a}
B.~Morris.
\newblock {F}ive{T}hirty{E}ight---{S}keptical {F}ootball: {P}atriots vs.
  {C}ardinals and an {I}nteractive {H}istory of the {N}{F}{L}, 2014.
\newblock
  \href{http://fivethirtyeight.com/features/skeptical-football-patriots-vs-cardinals-and-an-interactive-history-of-the-nfl/}{http://fivethirtyeight.com/features/skeptical-football-patriots-vs-cardinals-and-an-interactive-history-of-the-nfl/},
  accessed on June 25, 2015.

\bibitem{bryant2000a}
J.~Bryant and A.~A. Raney.
\newblock Sports on the screen.
\newblock In D.~Zillmann and P.~Vorderer, editors, {\em Media entertainment:
  {The} psychology of its appeal}, {LEA}'s communication series., pages
  153--174. Lawrence Erlbaum Associates Publishers, Mahwah, NJ, US, 2000.

\bibitem{pawlowski2013a}
T.~Pawlowski.
\newblock Testing the uncertainty of outcome hypothesis in {E}uropean
  professional football: {A} stated preference approach.
\newblock {\em Journal of Sports Economics}, 14(4):341--367, Aug. 2013.

\end{thebibliography}

\end{document}